\documentclass[aps,prc,superscriptaddress,twoside,twocolumn,nofootinbib,showpacs]{revtex4-1}
\usepackage{amsmath}
\usepackage{amssymb}
\usepackage{graphicx}
\usepackage{braket}
\begin{document}

\title{Explicit derivation of the completeness condition in pseudoscalar meson photoproduction}
\author{K. Nakayama}
\affiliation{Department of Physics and Astronomy, University of Georgia,
Athens, GA 30602, USA}

\begin{abstract}
By exploiting the underlying symmetries of the relative phases of the pseudoscalar meson photoproduction amplitude, 
we provide a consistent and explicit mathematical derivation of the completeness condition for the observables in this reaction. In particular, we determine all the possible sets of four double-spin observables that resolve the phase ambiguity of the amplitude in transversity basis up to an overall phase. The present work substantiates and corroborates the original findings 
of Chiang and Tabakin [Phys. Rev. C \textbf{55}, 2054 (1997)].  It is found, however, that the completeness condition of four double-spin observables to resolve the phase ambiguity holds only when the relative phases \textit{do not} meet the condition of equal magnitudes.
In situations where this condition occurs, it is shown that one needs extra chosen observables, resulting in the minimum number of  observables required to resolve the phase ambiguity reaching up to eight, depending on the particular set of four double-spin observables considered. 
Furthermore, a way of gauging when the condition of equal magnitudes occurs is provided.

\end{abstract}


\maketitle


\section{Introduction}

The issue of model-independent determination of the pseudoscalar meson photoroduction amplitude has attracted much attention since the early stage of investigation of this reaction process. In particular, early papers on the  minimum number of experimental observables required to determine the pseudoscalar meson photoproduction amplitude -- the so-called \textit{complete experiments} -- have resulted in contradictory findings (for a brief account on these, see Ref.~\cite{BDS75}). Barker, Donnachie and Storrow \cite{BDS75}
have cleared this situation, by deriving the necessary and sufficient conditions for determining the full photoproduction amplitude up to discrete ambiguities. They also provided the rules for choosing further measurements to resolve these ambiguities. According to these authors, for a given kinematics (total energy of the system and meson production angle), one requires nine observables 
to determine the full reaction amplitude up to an arbitrary overall phase.
Keaton and Workman \cite{KW96}, however, have realized that there are cases obeying the rules given in Ref.~\cite{BDS75} that still leave unsolved ambiguities. Finally, Chiang and Tabakin \cite{ChT97}, have shown that, instead of nine observables as claimed in Ref.~\cite{BDS75}, one requires a minimum of eight carefully chosen observables for a complete experiment. 
Apart from solving for the amplitude magnitudes and phases directly, Chiang and Tabakin \cite{ChT97} in their study, have also used a bilinear helicity product formulation to map an algebra of measurements over to the well-known algebra of the 4x4 gamma matrices. This latter method leads to an alternate proof that eight carefully chosen experiments suffice for determining the transversity amplitudes completely. 
The issue of complete experiments has been also discussed by Moravcsik \cite{Mo84} in the context of a general reaction process. There, a very similar approach to that of Ref.~\cite{ChT97} is used for resolving the discrete phase ambiguities of the reaction amplitude with a geometrical interpretation.
Sandorf \textit{et al.} \cite{SHKL11} have concluded among other things that, while a
mathematical solution to the problem of determining an amplitude free of ambiguities
may require eight observables \cite{ChT97}, experiments with realistically achievable uncertainties will require a significantly larger number of observables. 
Also, the Gent group has extended much effort along this line \cite{VRCV13,NVR15,NRIG16}.
Recently, with the advances in experimental techniques, many spin-observables in photoproduction reactions became possible to be measured and this has attracted much interest in constraints on partial-wave analysis in the context of complete experiments \cite{Workman11,WPBTSOK11,DMIM11,WBT14,WTWDH17,FGRGP18}. Of particular interest in this connection is the issue of whether the baryon resonances can be extracted model independently or with minimal model inputs. Efforts in this direction are currently in progress  \cite{WBT14,WTWDH17,FGRGP18}.

In this work, we revisit the problem of complete experiments in pseudoscalar meson photoproduction from a mathematical point of view, i.e., under ideal experiments with zero uncertainties. Thus, it is most directly related to the work of Ref.~\cite{ChT97}. We tackle this problem by solving for the amplitude magnitudes and phases directly, as has been done in Ref.~\cite{ChT97}. 
In doing so, we shall reveal and exploit the underlying symmetries of the relative phases of the photoproduction amplitude, which allows a consistent and explicit mathematical derivation of the completeness condition for the observables covering all the relevant cases.   
The completeness condition of a set of four double-spin observables to resolve the phase ambiguity of the transversity amplitude is shown to hold,  except in situations where the equal relative-phase magnitudes relation - 
as specified in Eq.(\ref{eq:3-0}) later in Sec.~\ref{sec:equalphase} - occur.
It will be shown that, when this situation occurs, one needs up to seven chosen double-spin observables, instead of four, to resolve the phase ambiguity.
Furthermore, in the particular situation where the relative phases vanish, eight chosen double-spin observables are required to resolve the phase ambiguity.

The paper is organized as follows. In Sec.~\ref{sec:notations}, we introduce the notations used throughout this work and express the observables as bilinear combinations of the four basic transversity amplitudes. In addition, we group the observables and classify them in cases which are convenient for determining the possible sets of four observables that resolve the phase ambiguity. 
In Secs.~\ref{sec:22},  \ref{sec:211} and \ref{sec:314}, we determine these sets of four double-spin observables, according to the classification introduced in Sec.~\ref{sec:notations}. There, we also consider the cases where the restriction on the relative phases for the completeness condition of the four observables is not satisfied. In Sec.~\ref{sec:equalphase}, we discuss how to identify when this restriction is violated. Finally, a summary is given in Sec.~\ref{sec:summary}.

\section{Notations} \label{sec:notations}

The basic four independent amplitudes, $M_j\ (j=1,\cdots, 4)$, that constitute the full pseudoscalar photoproduction amplitude can be expressed as
\begin{equation}
M_j  = r_j e^{i\phi_j} \ , \ \left\{
\begin{array}{lll}
r_j & = & {\rm magnitude} \ ,\\
\phi_j & = & {\rm phase} \ .
\end{array} \right.
\label{eq:2}
\end{equation}
Then, following Ref.~\cite{ChT97}, the 16 non-redundant observables can be expressed in terms of these amplitudes $M_j$ in transversity basis and grouped according to
\begin{equation}
{\cal S} = \left\{
\begin{array}{llrl} 
d\sigma / d\Omega  & =  \frac12 \left[ | r_1 |^2 + | r_2 |^2 + | r_3 |^2 + | r_4 |^2 \right] \ ,  \vspace{0.2cm}  \\ 
\Sigma  & =  \frac12 \left[ | r_1 |^2 + | r_2 |^2 - | r_3 |^2 - | r_4 |^2 \right] \ ,  \vspace{0.2cm}  \\ 
T  & =  \frac12 \left[ | r_1 |^2 - | r_2 |^2 - | r_3 |^2 + | r_4 |^2 \right] \ ,  \vspace{0.2cm}  \\ 
P  & =  \frac12 \left[ - | r_1 |^2 + | r_2 |^2 - | r_3 |^2 + | r_4 |^2 \right] \ ,  
\end{array} \right.
\label{eq:comb1}
\end{equation}
\begin{equation}
{\cal BT} = \left\{
\begin{array}{llrl} 
O^a_{1+} & \equiv & -G & =  B_{13} \sin\phi_{13} + B_{24} \sin\phi_{24} \ ,  \vspace{0.1cm}  \\ 
O^a_{1-}  & \equiv &  F & =  B_{13} \sin\phi_{13} -  B_{24} \sin\phi_{24} \ ,   \vspace{0.1cm} \\
O^a_{2+} & \equiv & E &  =  B_{13} \cos\phi_{13} + B_{24} \cos\phi_{24} \ ,   \vspace{0.1cm} \\ 
O^a_{2-}  & \equiv & H &  =  B_{13} \cos\phi_{13} -  B_{24} \cos\phi_{24} \ , 
\end{array} \right.
\label{eq:comb1a}
\end{equation}
\begin{equation}
{\cal BR} = \left\{
\begin{array}{llrl} 
O^b_{1+} & \equiv &   O_z & =  B_{14} \sin\phi_{14} + B_{23} \sin\phi_{23} \ ,   \vspace{0.1cm} \\
O^b_{1-}  & \equiv & - C_x & =  B_{14} \sin\phi_{14} -  B_{23} \sin\phi_{23} \ ,  \vspace{0.1cm} \\
O^b_{2+} & \equiv & - C_z & =  B_{14} \cos\phi_{14} + B_{23} \cos\phi_{23} \ ,  \vspace{0.1cm} \\
O^b_{2-}  & \equiv & - O_x & =  B_{14} \cos\phi_{14} -  B_{23} \cos\phi_{23} \ , 
\end{array} \right.
\label{eq:comb1b}
\end{equation}
\begin{equation}
{\cal TR} = \left\{
\begin{array}{llrl} 
O^c_{1+} & \equiv &  - L_x & =  B_{12} \sin\phi_{12} + B_{34} \sin\phi_{34} \ ,  \vspace{0.1cm}  \\
O^c_{1-}  & \equiv & - T_z & =  B_{12} \sin\phi_{12} -  B_{34} \sin\phi_{34} \ ,  \vspace{0.1cm} \\
O^c_{2+} & \equiv & - L_z & =  B_{12} \cos\phi_{12} + B_{34} \cos\phi_{34} \ ,  \vspace{0.1cm} \\ 
O^c_{2-}  & \equiv &   T_x & =  B_{12} \cos\phi_{12} -  B_{34} \cos\phi_{34} \ ,  
\end{array} \right.
\label{eq:comb1c}
\end{equation}
where
\begin{equation}
B_{ij}  \equiv r_i r_j  \qquad {\rm and} \qquad 
\phi_{ij}  \equiv \phi_i - \phi_j \  .
\label{eq:notation0}
\end{equation}

In the following we refer to $\phi_{ij}$ as the relative phase.
The observables in ${\cal S}$ include the unpolarized cross section, $d\sigma/d\Omega$, and single-spin observables $\Sigma$ (beam asymmetry), $T$ (target asymmetry) and $P$ (recoil asymmetry). It is clear from Eq.(\ref{eq:comb1}) that, together, they determine uniquely the magnitudes of the basic four amplitudes in transversity basis.   
 Throughout this work, these four observables are assumed to be measured, so that the magnitudes of the basic transversity amplitudes are known. 
The remaining observables given in Eqs.(\ref{eq:comb1a},\ref{eq:comb1b},\ref{eq:comb1c}) are all double-spin observables and some combinations of them will serve to determine the phases of the four transversity amplitudes up to an overall phase,  i.e., the three relative phases $\phi_{ij}$ involved.  
We refer to the observables in each of ${\cal BT}$ (beam-target asymmetry), ${\cal BR}$ (beam-recoil asymmetry)  and ${\cal TR}$ (target-recoil asymmetry)  as a group.  We use $a = {\cal BT}$, $b = {\cal BR}$ and $c={\cal TR}$. 

In Ref.~\cite{ChT97}, the unnormalized spin asymmetries are denoted by $\check{\Omega}^\beta$, i.e., $\check{\Omega}^\beta \equiv (d\sigma/ d\Omega) \Omega^\beta$, where $\Omega^\beta$ stands for a given spin asymmetry specified by the index $\beta$.  Throughout this work, we simply use the same notation $\Omega^\beta$ for the unnormalized spin asymmetries ($(d\sigma/ d\Omega) \Omega^\beta \to \Omega^\beta$) to avoid overloading the notations.  For example, $\Sigma$ in Eq.(\ref{eq:comb1}) actually stands for $(d\sigma/ d\Omega) \Sigma$, and so on.

\vskip 0.5cm
From the above list of observables, one sees that all possible sets of four double-spin observables can be  obtained by considering the following cases:
\begin{itemize}

\item[1)]   \textbf{($2 + 2$) case:} \ 
two pairs of observables, each pair from distinct groups.

\item[2)]  \textbf{($2 + 1 + 1$) case:} \ 
a pair of observables from one group and two other observables, one from each of the remaining two groups. 

\item[3)]   \textbf{($3 + 1$) case:} \ 
three observables from one group and one observable from another group.

\item[4)]   \textbf{$4$ case:} \ 
all four observables from one group.

\end{itemize}

In the following we shall consider each of the cases listed above.

\section{Phase fixing for the $2 + 2$ case} \label{sec:22}

We start by noticing that there are two basic types of combination of a pair of observables ($O^m_{n\nu}, O^m_{n'\nu'}$) in a given group, one type with $n=n'$ and the other with $n \ne n'$. Here, ($m=a,b,c$), \ ($n,n'=1,2$) and ($\nu,\nu'= \pm$). A pair of observables of the type ($O^m_{n+}, O^m_{n -}$) leads to a four-fold phase ambiguity, with two-fold ambiguity in each of the relative phases involved, $\phi_{ij}$ and $\phi_{kl}$. There are two distinct pairs of this type ($n=1,2$) in each group.
On the other hand, a pair of observables of the type ($O^m_{1\nu}, O^m_{2\nu'}$), leads only to a two-fold phase ambiguity. We have four distinct pairs of this type $(\nu,\nu'= \pm)$ in each group. 

\vskip 0.5cm
To see the properties mentioned above, let us consider all the possible pairs one can form  in a given group, say, group $a = {\cal BT}$.
For the pair  ($O^a_{1+}, O^a_{1-}) =(-G, F)$, we have from Eq.(\ref{eq:comb1a}),
\begin{align}
O^a_{1+} & = B_{13} \sin\phi_{13} + B_{24} \sin\phi_{24}  \ ,   \nonumber \\
O^a_{1-} & = B_{13} \sin\phi_{13} - B_{24} \sin\phi_{24}  \  ,
\label{eq:2-1}
\end{align}
which leads to
\begin{align}
\sin\phi_{13}  & = \frac{O^a_{1+} + O^a_{1-}}{2 B_{13}}  \Longrightarrow  \phi_{13} = \left\{
\begin{array}{l}
\alpha_{13}  \ ,   \\
\pi - \alpha_{13}  \ , 
\end{array} \right. 
\nonumber \\ \nonumber \\
\sin\phi_{24}  & = \frac{O^a_{1+} - O^a_{1-}}{2 B_{24}}  \Longrightarrow  \phi_{24} = \left\{
\begin{array}{l}
\alpha_{24}  \ , \\ 
\pi - \alpha_{24}  \ , 
\end{array} \right.  
\nonumber \\
\label{eq:2-2}
\end{align}
where $-\pi/2 \le \alpha_{13}, \alpha_{24}  \le +\pi/2$; $\alpha_{ij}$'s are uniquely defined. In the following, we use the notation $\phi_{ij}^\lambda$ to designate
\begin{equation}
\phi_{ij}^+ = \alpha_{ij} \ , \qquad\qquad  \phi_{ij}^- = \pi - \alpha_{ij} \ .
\label{eq:2-3}
\end{equation}
Note that a (relative) phase is meaningful only  \textit{modulo} $2\pi$.

\vskip 0.5cm
Analogously, for the pair  ($O^a_{2+}, O^a_{2-}) =(E, H)$, we have from Eq.(\ref{eq:comb1a}),
\begin{align}
O^a_{2+} & = B_{13} \cos\phi_{13} + B_{24} \cos\phi_{24}  \ ,   \nonumber \\
O^a_{2-} & = B_{13} \cos\phi_{13} - B_{24} \cos\phi_{24}  \  ,
\label{eq:2-4}
\end{align}
which leads to the two-fold ambiguity 
\begin{equation}
\phi_{ij}^+ = \alpha_{ij} \ , \qquad\qquad  \phi_{ij}^- = - \alpha_{ij} \ ,
\label{eq:2-6}
\end{equation}
where $\alpha_{ij}$ is uniquely defined with $0 \le \alpha_{ij} \le \pi$.  

\vskip 0.5cm
Next we consider the pair  ($O^a_{1+}, O^a_{2-}) =(-G, H)$. From Eq.(\ref{eq:comb1a}),
\begin{align}
O^a_{1+} & = B_{13} \sin\phi_{13} + B_{24} \sin\phi_{24}  \ ,   \nonumber \\
O^a_{2-} & = B_{13} \cos\phi_{13} - B_{24} \cos\phi_{24}  \ .
\label{eq:2-7}
\end{align}
We first combine the above two expressions into
\begin{equation}
O^a_{1+}{^2} + O^a_{2-}{^2} = B^2_{13} + B^2_{24} - 2 B_{13} B_{24} \cos(\phi_{13} + \phi_{24})  \ .
\label{eq:2-8}
\end{equation}
Now, we define angle $\zeta \equiv \zeta^m_{n\nu, n'\nu'}$ through
\footnote{$\zeta^m_{n\nu, n'\nu'}$ has a geometrical interpretation as the polar angle of a
vector in a 2-dimensional coordinate system, where $O^m_{n\nu}$ defines the $x$-coordinate and $O^m_{
n'\nu'}$, the $y$-coordinate. This provides an intuitive understanding of the fact that such an angle, 
$\zeta^m_{n\nu, n'\nu'}$, can indeed always be found.}
\begin{equation}
\cos\zeta \equiv \frac{O^m_{n\nu}}{N}  \ , \quad\qquad \sin\zeta \equiv \frac{O^m_{n'\nu'}}{N} \ , 
\label{eq:2-8a}
\end{equation}
with $N \equiv  N^m_{n\nu, n'\nu'} \equiv \sqrt{O^m_{n\nu}{^2} + O^m_{n'\nu'}{^2}}$.
In the following we simply use $\zeta$ and $N$ to avoid the heavy notation, but it should be kept in mind that they depend on the given pair of observables. 
For the pair under consideration, we have
\begin{equation}
\cos\zeta \equiv \frac{O^a_{1+}}{N}  \ , \quad\qquad \sin\zeta \equiv \frac{O^a_{2-}}{N} \ , 
\label{eq:2-8a1}
\end{equation}
with $N \equiv \sqrt{O^a_{1+}{^2} + O^a_{2-}{^2}}$.

\vskip 0.3cm
Then, Eq.(\ref{eq:2-7}) can be expressed in terms of $\zeta$ as
\begin{align}
N \cos\zeta & = B_{13} \sin\phi_{13} + B_{24} \sin\phi_{24}  \ ,   \nonumber \\
N \sin\zeta & = B_{13} \cos\phi_{13} - B_{24} \cos\phi_{24}  \ .
\label{eq:2-9}
\end{align}
Multiplying the first equality in the above equation by $\sin\phi_{24}$ and the second one by $\cos\phi_{24}$ and subtracting the second from the first, we arrive at
\begin{equation}
\cos(\phi_{13} + \phi_{24}) = \frac{ B_{24} + N  \sin(\zeta - \phi_{24}) }{B_{13} } \ .
\label{eq:2-10}
\end{equation}
Inserting the above result into Eq.(\ref{eq:2-8}) yields
\begin{equation}
\sin( \zeta - \phi_{24}) = \frac{B_{13}^2 - B_{24}^2 - N^2}{2 N B_{24}} \ ,
\label{eq:2-11}
\end{equation}
leading to the following  two-fold ambiguity for $\phi_{24}$: 
\begin{equation}
\phi_{24} = \left\{
\begin{array}{l}
\zeta - \alpha_{24} \ , \\
\zeta - \pi + \alpha_{24} \ . 
\end{array} \right.
\label{eq:2-12}
\end{equation}

Analogously, from Eqs.(\ref{eq:2-8},\ref{eq:2-9}), we find that
\begin{equation}
\sin(\zeta + \phi_{13}) = \frac{B_{13}^2 - B_{24}^2 + N^2}{2 N B_{13}} \ ,
\label{eq:2-13}
\end{equation}
leading to the two-fold ambiguity 
\begin{equation}
\phi_{13} = \left\{
\begin{array}{l}
- \zeta + \alpha_{13} \ , \\
- \zeta + \pi - \alpha_{13} \ .
\end{array} \right.
\label{eq:2-14}
\end{equation}
Note that, in Eqs.(\ref{eq:2-12},\ref{eq:2-14}), phases $\alpha_{24}$ and $\alpha_{13}$  are uniquely defined by
\begin{align}
\sin( \alpha_{24}) & = \frac{B_{13}^2 - B_{24}^2 - N^2}{2 N B_{24}}  \ , \nonumber \\  
\sin( \alpha_{13}) & = \frac{B_{13}^2 - B_{24}^2 + N^2}{2 N B_{13}} \ , 
\label{eq:2-5}
\end{align}
with $-\pi/2 \le \alpha_{13}, \alpha_{24}  \le +\pi/2$.

\vskip 0.3cm
Equations (\ref{eq:2-12},\ref{eq:2-14}) show that $\phi_{13}$ and $\phi_{24}$ have a two-fold ambiguity each. However, there is another constraint that $\cos(\phi_{13}+\phi_{24})$ is uniquely defined by Eq.(\ref{eq:2-8}). 
Then, first we note that the sum of $\phi_{13}$ and $\phi_{24}$ should be of the form $\phi_{13}+\phi_{24} = \pm \tilde{\alpha}$.
Combining this with Eqs.(\ref{eq:2-12},\ref{eq:2-14}), it leads to the following possibilities for $\tilde{\alpha}$:
\begin{equation}
\tilde{\alpha} = \left\{
\begin{array}{l}
\lambda\left(\phi^\lambda_{13} + \phi^\lambda_{24} \right) = \left(\alpha_{13} - \alpha_{24}\right)\ , \\
\lambda \left(\phi^\lambda_{13} + \phi^{\lambda'}_{24} \right) = \left(\alpha_{13} + \alpha_{24} - \pi\right)\ \ ,
\end{array}\right.
\label{eq:2-15b}
\end{equation}
where the notation introduced in Eq.(\ref{eq:2-3}) has been used. Here, $\lambda,\lambda'=\pm$ and $\lambda \ne \lambda'$.

Next, we calculate $\cos(\phi_{13}+\phi_{24})= \cos(\pm \tilde{\alpha})$, with $\tilde{\alpha}$ given in Eq.~(\ref{eq:2-15b}). For $\tilde{\alpha} = \alpha_{13} - \alpha_{24}$, we obtain
\begin{align}
\cos(\phi_{13}+ \phi_{24}) & =  \cos(\pm(\alpha_{13}-\alpha_{24})) \nonumber \\
& = \cos\alpha_{13}\cos\alpha_{24} + \sin\alpha_{13}\sin\alpha_{24}  \nonumber \\
 & = \sqrt{(1- \sin^2\alpha_{13})(1 - \sin^2\alpha_{24})} \nonumber \\
 &\quad  + \sin\alpha_{13}\sin\alpha_{24}  \nonumber \\
 & =  \frac{B_{13}^2 + B_{24}^2 - N^2}{2B_{13}B_{24}} \ ,
 \label{eq:2-15aa}
 \end{align}
where Eq.(\ref{eq:2-5}) has been used. This result coincides with Eq.(\ref{eq:2-8}).
For $\tilde{\alpha} = \alpha_{13}+\alpha_{24}- \pi$, on the other hand, it is immediately seen that the result 
for $\cos(\phi_{13}+\phi_{24})$ does not agree with Eq.(\ref{eq:2-8}) since, in this case, apart from an overall 
sign, all that changes from the $\tilde{\alpha} = \alpha_{13}-\alpha_{24}$ case is the change in the sign 
of the term $\sin\alpha_{13}\sin\alpha_{24}$ - which is non-zero in general - in Eq.~(\ref{eq:2-15aa}).

Thus, we conclude that Eq.(\ref{eq:2-8}), together with  Eqs.(\ref{eq:2-12},\ref{eq:2-14}), leads to
\begin{equation}
\phi_{13} + \phi_{24} = \pm (\alpha_{13} - \alpha_{24}) \ ,
\label{eq:2-15}
\end{equation}
i.e., we end up with only two-fold ambiguity for $\phi_{13}$ and $\phi_{24}$, viz.,
\begin{equation}
\left\{
\begin{array}{ll} 
\phi_{13} = & - \zeta + \alpha_{13} \ ,  \\
\phi_{24} = &\ \  \,  \zeta - \alpha_{24} \ ,
\end{array} \right.
\quad {\rm or} \quad
\left\{
\begin{array}{ll} 
\phi_{13} = &  - \zeta - \alpha_{13} + \pi \ , \\
 \phi_{24} = & \ \ \,  \zeta + \alpha_{24} - \pi \ .
 \end{array} \right.
\label{eq:2-16}
\end{equation}

\vskip 0.5cm
For the pair $(O^a_{1-}, O^a_{2-}) = (F, H)$, 
\begin{align}
O^a_{1-} & = B_{13} \sin\phi_{13} - B_{24} \sin\phi_{24}  \ ,   \nonumber \\
O^a_{2-} & = B_{13} \cos\phi_{13} - B_{24} \cos\phi_{24}  \ ,
\label{eq:2-19}
\end{align}
the results can be readily obtained by simply changing the sign of $\phi_{24}$ everywhere in the results of the previous case of  $(O^a_{1+}, O^a_{2-})$. We obtain
\begin{equation}
\left\{
\begin{array}{ll} 
\phi_{13} = & - \zeta + \alpha_{13} \ ,  \\
\phi_{24} = & - \zeta + \alpha_{24} \ ,
\end{array} \right.
\quad {\rm or} \quad
\left\{
\begin{array}{ll} 
\phi_{13} = &  - \zeta - \alpha_{13} + \pi \ , \\
 \phi_{24} = & - \zeta - \alpha_{24} + \pi \ .
 \end{array} \right.
\label{eq:2-18a}
\end{equation}

\vskip 0.5cm
For the pair $(O^a_{1-}, O^a_{2+}) = (F, E)$,
\begin{align}
O^a_{1-} & = B_{13} \sin\phi_{13} - B_{24} \sin\phi_{24}  \ ,   \nonumber \\
O^a_{2+} & = B_{13} \cos\phi_{13} + B_{24} \cos\phi_{24}  \ ,
\label{eq:2-17}
\end{align}
the only change from the previous case of $(O^a_{1+}, O^a_{2-})$, is in the sign of $B_{24}$.
Thus, we can simply follow the steps of the derivation for the case of $(O^a_{1+}, O^a_{2-})$,
making there the replacement $B_{24} \to - B_{24}$. 
This leads to the change in the constraint given by Eq.(\ref{eq:2-15}) to
\begin{equation}
\phi_{13} + \phi_{24} = \pm (\alpha_{13} - \alpha_{24} + \pi) \ .
\label{eq:2-15a}
\end{equation}
Thus, we obtain the two-fold ambiguity
\begin{equation}
 \left\{
\begin{array}{ll} 
\phi_{13} = &  - \zeta - \alpha_{13} + \pi \ , \\
\phi_{24} = &\ \  \,  \zeta + \alpha_{24} \ ,
\end{array} \right.
\quad {\rm or} \quad
 \left\{
\begin{array}{ll} 
\phi_{13} = & - \zeta + \alpha_{13} \ ,  \\
 \phi_{24} = & \ \ \,  \zeta - \alpha_{24} + \pi \ .
 \end{array} \right.
\label{eq:2-16a}
\end{equation}

\vskip 0.5cm
For the pair $(O^a_{1+}, O^a_{2+}) = (-G, E)$,
\begin{align}
O^a_{1+} & = B_{13} \sin\phi_{13} + B_{24} \sin\phi_{24}  \ ,   \nonumber \\
O^a_{2+} & = B_{13} \cos\phi_{13} + B_{24} \cos\phi_{24}  \ ,
\label{eq:2-17aa}
\end{align}
we simply flip the sign of $\phi_{24}$ in Eq.(\ref{eq:2-16a}). We have 
\begin{equation}
\left\{
\begin{array}{ll} 
\phi_{13} = & - \zeta - \alpha_{13} + \pi \ ,  \\
\phi_{24} = & - \zeta - \alpha_{24} \ ,
\end{array} \right.
\quad {\rm or} \quad
\left\{
\begin{array}{ll} 
\phi_{13} = &  - \zeta + \alpha_{13}  \ , \\
 \phi_{24} = & - \zeta + \alpha_{24} - \pi \ . 
 \end{array} \right. \\ 
\label{eq:2-18}
\end{equation}

To avoid any confusion, we emphasize that,  in all the cases discussed above, $(O^a_{1\pm}, O^a_{2\pm})$ (with the signs $\pm$ being independent), the phases $\alpha_{13}$ and $\alpha_{24}$ 
are uniquely defined and given by Eq.(\ref{eq:2-5}).

\vskip 0.5cm
From the preceding considerations in this section, we conclude that 
\begin{itemize}

\item[i)]
Any pair of observables of the form ($O^m_{1+}\,, O^m_{1-}$) leads to a four-fold phase ambiguity of the form given by Eq.(\ref{eq:2-3}), while any pair of the form ($O^m_{2+}\,, O^m_{2-}$) leads to a four-fold ambiguity of the form given by Eq.(\ref{eq:2-6}). These result in
(in view of the consistency relations given by Eq.(\ref{eq:2-20}) that shall be used later on to help resolve the phase ambiguity)   
\begin{equation}
\qquad\, (O^a_{1+} , O^a_{1-}) :   \left\{ 
\begin{array}{l} 
 \phi^+_{13} - \phi^+_{24}  = \ \ \,  (\alpha_{13} - \alpha_{24}) \ , \\ 
 \phi^+_{13} - \phi^-_{24}  = \ \ \, [(\alpha_{13} + \alpha_{24}) - \pi] \ , \\ 
 \phi^-_{13} - \phi^+_{24}  = - [(\alpha_{13} + \alpha_{24}) - \pi] \ , \\ 
 \phi^-_{13} - \phi^-_{24}  =  - (\alpha_{13} - \alpha_{24}) \ ,  \\
 \\ 
 \phi^+_{13} + \phi^+_{24}  = \ \ \,  (\alpha_{13} + \alpha_{24}) \ , \\ 
 \phi^+_{13} + \phi^-_{24}  = \ \ \, (\alpha_{13} - \alpha_{24}) + \pi \ , \\ 
 \phi^-_{13} + \phi^+_{24}  = - (\alpha_{13} - \alpha_{24}) + \pi \ , \\ 
 \phi^-_{13} + \phi^-_{24}  =  - (\alpha_{13} + \alpha_{24}) \ , 
\end{array}   \right. 
\label{eq:a1a1-phase}
\end{equation}
and
\begin{equation}
\ \ \ \ \ \ \ \   (O^a_{2+} , O^a_{2-}) :  \left\{
\begin{array}{l} 
 \phi^+_{13} - \phi^+_{24}  = \ \ \, (\alpha_{13} - \alpha_{24})  \ ,   \\
 \phi^+_{13} - \phi^-_{24}  = \ \ \, (\alpha_{13} + \alpha_{24})  \ ,   \\
 \phi^-_{13} - \phi^+_{24}  =  - (\alpha_{13} + \alpha_{24})  \ ,   \\
 \phi^-_{13} - \phi^-_{24}  =  - (\alpha_{13} - \alpha_{24})  \ ,   \\
\\
 \phi^+_{13} + \phi^+_{24}  = \ \ \,  (\alpha_{13} +\alpha_{24})  \ ,   \\
 \phi^+_{13} + \phi^-_{24}  = \ \ \, (\alpha_{13} - \alpha_{24})  \ ,   \\
 \phi^-_{13} + \phi^+_{24}  =  - (\alpha_{13} - \alpha_{24})  \ ,   \\
 \phi^-_{13} + \phi^-_{24}  =  - (\alpha_{13} + \alpha_{24})  \ ,   \\
\end{array} \right.
\label{eq:a2a2-phase}
\end{equation}

\item[ii)]
Any pair of observables of the form $(O^m_{1\pm}, O^m_{2\mp}) = (O^m_{1+}, O^m_{2-})\ {\rm or}\ (O^m_{1-}, O^m_{2+})$, leads to a two-fold ambiguity of the form given by Eqs.(\ref{eq:2-16},\ref{eq:2-16a}), while any pair of the form ($O^m_{1\nu}, O^m_{2\nu}$), leads to a two-fold ambiguity of the form given by Eqs.(\ref{eq:2-18a},\ref{eq:2-18}). These result in (recall that (relative) phases are \textit{modulo} $2\pi$)
\begin{equation}
\ \ \ \ \ \ \ \   (O^a_{1-} , O^a_{2-}) :  \left\{
\begin{array}{l} 
 \phi^\lambda_{13} - \phi^\lambda_{24}  = \lambda (\alpha_{13} - \alpha_{24})  \ ,  \\
 \phi^\lambda_{13} + \phi^\lambda_{24}  = -2\zeta + \lambda (\alpha_{13} +\alpha_{24})  \ ,   \\
\end{array} \right.
\label{eq:a1ma2m-phase}
\end{equation}
\begin{equation}
\ \ \ \ \ \ \ \   (O^a_{1+} , O^a_{2-}) :  \left\{
\begin{array}{l} 
 \phi^\lambda_{13} - \phi^\lambda_{24}  = -2\zeta + \lambda (\alpha_{13} +\alpha_{24})  \ ,   \\
 \phi^\lambda_{13} + \phi^\lambda_{24}  = \lambda (\alpha_{13} - \alpha_{24})  \ ,  \\
\end{array} \right.
\label{eq:a1pa2m-phase}
\end{equation}
with $\lambda=\pm$, and
\begin{equation}
\ \ \ \ \ \ \ \   (O^a_{1+} , O^a_{2+}) :  \left\{
\begin{array}{l} 
 \phi^+_{13} - \phi^-_{24}  = \ \ \, (\alpha_{13} - \alpha_{24}) + \pi   \ ,   \\
 \phi^-_{13} - \phi^+_{24}  =  - (\alpha_{13} - \alpha_{24}) + \pi \ ,   \\
\\
 \phi^+_{13} + \phi^-_{24}  = -2\zeta + (\alpha_{13} + \alpha_{24}) - \pi \ ,   \\
 \phi^-_{13} + \phi^+_{24}  = -2\zeta  - (\alpha_{13} + \alpha_{24})  + \pi \ ,   \\
\end{array} \right.
\label{eq:a1pa2p-phase}
\end{equation}
\begin{equation}
\ \ \ \ \ \ \ \   (O^a_{1-} , O^a_{2+}) :  \left\{
\begin{array}{l} 
 \phi^+_{13} - \phi^-_{24}  = -2\zeta + (\alpha_{13} + \alpha_{24}) - \pi \ ,   \\
 \phi^-_{13} - \phi^+_{24}  = -2\zeta  - (\alpha_{13} + \alpha_{24})  + \pi \ ,   \\
\\
 \phi^+_{13} + \phi^-_{24}  =  - (\alpha_{13} - \alpha_{24}) + \pi \ ,   \\
 \phi^-_{13} + \phi^+_{24}  = \ \ \, (\alpha_{13} - \alpha_{24}) + \pi   \ .   \\
\end{array} \right.
\label{eq:a1ma2p-phase}
\end{equation}
\end{itemize}

From the results obtained above for the pairs of observables $(O^a_{n\nu}, O^a_{n'\nu'})\ (n,n'=1,2\ {\rm and}\  \nu,\nu'=\pm$ with $(n\nu) \ne (n'\nu'))$  in group $a= {\cal BT}$ (cf.~Eq.(\ref{eq:comb1a})), it is straightforward to obtain the corresponding results for the pairs of observables in other two groups $b={\cal BR}$ and $c={\cal TR}$ (cf.~Eqs.(\ref{eq:comb1b},\ref{eq:comb1c})). All we have to do is to replace $(O^a_{n\nu}, O^a_{n'\nu'})$ by $(O^m_{n\nu}, O^m_{n'\nu'})\ (m=b,c)$ and the relative phases $\phi_{13}$ and $\phi_{24}$, respectively, by $\phi_{14}$ and $\phi_{23}$ in the case $m=b$ or by $\phi_{12}$ and $\phi_{34}$ in the case $m=c$.

 \vskip 0.5cm
The discrete ambiguities exhibited by the relative phases so far in this section (cf.~Eqs.(\ref{eq:a1a1-phase},\ref{eq:a2a2-phase},\ref{eq:a1ma2m-phase},\ref{eq:a1pa2m-phase},\ref{eq:a1pa2p-phase},\ref{eq:a1ma2p-phase})) cannot be resolved without further constraint.  This is provided by the property obeyed by the relative phases ($\phi_{ij} \equiv \phi_i - \phi_j$): 
\footnote{Equation (\ref{eq:2-20a}) may be seen as a direct consequence of the fact that a complex number can be represented by a vector in the complex plane and that the sum of all angles between neighboring vectors in a given set of vectors is $2\pi$ (or zero since phases are \textit{modulo} $2\pi$).}
\begin{equation}
\phi_{12} + \phi_{23} + \phi_{34} = \phi_{14} \ .
\label{eq:2-20a}
\end{equation}
Here, it should be emphasized that this relation is satisfied up to an addition of multiples of $2\pi$, because phases are meaningful only modulo $2\pi$. 
 We refer to the above relation as the \textit{consistency} relation, because it is going to be used to check on the 'consistency' among the relative phases with discrete ambiguities as we have shown in our considerations up to this point. As the reader shall see, the consistency relation allows us to resolve the discrete
 ambiguities for certain sets of four chosen observables. Equation (\ref{eq:2-20a}) can be rewritten as 
\begin{subequations}
\begin{align}
\phi_{24} - \phi_{13} & = \phi_{34} - \phi_{12}    \qquad (a \longleftrightarrow c)  \ ,  \\
\phi_{24} + \phi_{13} & = \phi_{14} + \phi_{23}  \qquad (a \longleftrightarrow b) \ , \\
\phi_{34} + \phi_{12} & = \phi_{14} - \phi_{23}  \qquad (c \longleftrightarrow b)\ .
\end{align}
\label{eq:2-20}
\end{subequations}
The first relation in the above equation is used to relate the observables in group $a={\cal BT}$ to those in group $c={\cal TR}$, while the second relation connects the observables in group $a$ to those in group $b={\cal BR}$. The third relation connects the observables in group $b$ to those in group $c$.
Note that, apart from an irrelevant overall factor, Eq.(\ref{eq:2-20a}) leads to a unique relation which connects the relative phases belonging to two specific groups of observables as exhibited in Eq.(\ref{eq:2-20}). 
Equation (\ref{eq:2-20}) has been also used by the authors of Refs.~\cite{Mo84,ChT97} in their analyses.

\vskip 0.5cm
The logic for determining whether a given set of four observables can or cannot resolve the phase ambiguity is as follows. From the chosen set of four observables, using the appropriate consistency relation in Eq.(\ref{eq:2-20}), form all possible solutions due to the discrete ambiguities of the relative phases which, for the \textbf{($2+2$) case}, are given by Eqs.(\ref{eq:a1a1-phase},\ref{eq:a2a2-phase},\ref{eq:a1ma2m-phase},\ref{eq:a1pa2m-phase},\ref{eq:a1pa2p-phase},\ref{eq:a1ma2p-phase})). Then, check if these solutions are linearly independent (non-degenerated) or dependent (degenerated). If there is no degeneracy in the possible solutions (i.e., they are \textit{all} linearly independent), then, only one of them will be satisfied, in general, once the set of unique values of the phases $\alpha_{ij}$'s and $\zeta's\ (=\zeta^m_{n\nu, n'\nu'})$ is provided by the measurements of the  four observables in consideration.
\footnote{Recall that the unpolarized cross section and single-spin observables are assumed to be measured. They fix the magnitudes of the four basic transversity amplitudes which enter in the determination of $\alpha_{ij}$'s (cf.~Eq.(\ref{eq:2-5})). }
The precise relation of each $\alpha_{ij}$ to the corresponding $\phi_{ij}$ is known once the correct solution among the possible solutions is identified, thus, resolving the ambiguity of $\phi_{ij}$.  Hence, this set of four observables resolves the phase ambiguity. If the degeneracy occurs among the possible solutions, then, this set of observables  cannot resolve the ambiguity.  The logic just described applies to all \textbf{cases ($1, 2, 3, 4$)} specified at the end of the previous section.  Only the discrete ambiguities of the relative phases are case-dependent, as shown later in Secs.~\ref{sec:211},\ref{sec:314}.

It should be clear from the above consideration that, whether a set of four observables resolves the phase ambiguity or not, rests on the linear independence of the possible solutions provided by the consistency relation (cf.~Eqs.(\ref{eq:2-20})) for that set of four observables.

\vskip 1cm
We are now prepared to identify the possible sets of four double-spin observables that resolve the phase ambiguity of the transversity  amplitude in the \textbf{($2 + 2$) case} defined in item (1) of the preceding section. There are three basic combinations of the pairs of observables to be considered: 
\begin{itemize}

\item[aa)]
two pairs from item (i) above  with $4 \times 4 = 16$-fold phase ambiguity : $(O^m_{n+}\,, O^m_{n-})$ and $(O^{m'}_{n' +}\,, O^{m'}_{n' -})$ with $m \ne m'$.

\item[bb)]
two pairs from item (ii) above with $2 \times 2 = 4$-fold phase ambiguity : 
$(O^m_{1\nu}\,, O^m_{2\nu'})$ and $(O^{m'}_{1\mu}\,, O^{m'}_{2\mu'})$ with $m \ne m'$.   

\item[ab)]
one pair from item (i) and one pair from item (ii) with $4 \times 2 = 8$-fold phase ambiguity :
$(O^m_{n +}\,, O^m_{n -})$ and $(O^{m'}_{1\mu}\,, O^{m'}_{2\mu'})$ with $m \ne m'$.

\end{itemize}

\subsection{Case (aa)}

First, consider case (aa). To be concrete, choose the set of pairs $[(O^a_{2+}, O^a_{2-}) , (O^c_{2+}, O^c_{2-})]$. From Eqs.(\ref{eq:comb1a},\ref{eq:comb1b}), the observables in group $a$ contain relative phases $\phi_{13}$ and $\phi_{24}$, while those in group $c$ contain relative phases  $\phi_{12}$ and $\phi_{34}$. Then, using Eq.(\ref{eq:2-20}a), we have
\begin{equation}
\phi^\lambda_{13} - \phi^{\lambda'}_{24}  = \phi^{\lambda''}_{12} - \phi^{\lambda'''}_{34} \ , 
\label{eq:2-21}
\end{equation}
where the indices on which these relative phases depend have been written explicitly.
Inserting the corresponding four-fold phase ambiguity given by Eq.(\ref{eq:a2a2-phase}) into the above relation, we end up with 16 possible solutions 
\begin{align}
\pm \alpha_{13} \pm \alpha_{24}  & = \pm \alpha_{12} \pm \alpha_{34} \ , 
\label{eq:2-22}
\end{align}
where all four signs $\pm$ are independent. The 16 possible solutions given above are not all linearly independent. For example, consider the solution $\alpha_{13} + \alpha_{24} = \alpha_{12} + \alpha_{34}$ corresponding to $(\lambda, \lambda', \lambda'', \lambda''') = (+,-,+,-)$ in Eq.(\ref{eq:2-21}). This solution is degenerated with the solution 
$-(\alpha_{13} + \alpha_{24}) = -(\alpha_{12} + \alpha_{34})$ corresponding to $(\lambda, \lambda', \lambda'', \lambda''') = (-,+,-,+)$. Hence, the phase ambiguity cannot be resolved in this case.
It is also straightforward to see that none of the other combinations of the pairs of observables in case (aa)
resolve the ambiguity. This includes the corresponding sets of pairs of observables from group $a$ and group $b$ and from $b$ and $c$, in which cases we use the consistency relations given by Eqs.(\ref{eq:2-20}b) and (\ref{eq:2-20}c), respectively.

\subsection{Case (bb)}
For case (bb), let's start by considering the set of two pairs  $[(O^a_{1+}, O^a_{2-}) , (O^c_{1-}, O^c_{2-})]$.
From Eqs.(\ref{eq:comb1a},\ref{eq:comb1b}), the relative phases  involved for this combination are ($\phi_{13}, \phi_{24}$) and ($\phi_{12}, \phi_{34}$). Then, inserting Eqs.(\ref{eq:a1pa2m-phase},\ref{eq:a1ma2m-phase}) into Eq.(\ref{eq:2-20}a), yields
the following four possible solutions:
\begin{align}
-2 \zeta + (\alpha_{13} + \alpha_{24}) & =  \ \ \, (\alpha_{12} - \alpha_{34})  \ , \nonumber \\
-2 \zeta + (\alpha_{13} + \alpha_{24}) & =  - (\alpha_{12} - \alpha_{34})  \ , \nonumber \\
-2 \zeta -  (\alpha_{13} + \alpha_{24}) & = \ \ \, (\alpha_{12} - \alpha_{34}) \ , \nonumber \\
-2 \zeta -  (\alpha_{13} + \alpha_{24}) & = - (\alpha_{12} - \alpha_{34}) \ . 
\label{eq:2-26}
\end{align}
Since the above possible solutions are all linearly independent, there will be only one solution satisfied, in general, for the set of unique values of $\alpha_{13}, \alpha_{24}, \alpha_{12}, \alpha_{34}$ and $\zeta(=\zeta^a_{1+, 2-})$, once they are extracted from the measurements of the four observables in question. The correct solution, then, will tell us  the exact relation of each $\alpha_{ij}\ (ij=13, 24, 12, 34)$ to the corresponding $\phi_{ij}$, resolving the ambiguity of $\phi_{ij}$. Hence this set
of four observables will resolve the phase ambiguity. 

Consider now the set of pairs $[(O^a_{1+}, O^a_{2-}) , (O^c_{1+}, O^c_{2-})]$. 
Again, with the help of Eq.(\ref{eq:a1pa2m-phase}), Eq.(\ref{eq:2-20}a) leads to
\begin{align}
-2 \zeta + (\alpha_{13} + \alpha_{24}) & =  -2 \zeta' + (\alpha_{12} + \alpha_{34})  \ , \nonumber \\
-2 \zeta + (\alpha_{13} + \alpha_{24}) & =  -2 \zeta' -  (\alpha_{12} + \alpha_{34})  \ , \nonumber \\
-2 \zeta -  (\alpha_{13} + \alpha_{24}) & = -2 \zeta'  + (\alpha_{12} + \alpha_{34}) \ , \nonumber \\
-2 \zeta -  (\alpha_{13} + \alpha_{24}) & = -2 \zeta' - (\alpha_{12} + \alpha_{34}) \ . 
\label{eq:2-27}
\end{align}
Note that $\zeta$ is distinct from $\zeta'$ (cf. Eq.(\ref{eq:2-8a})). 
As in the previous case just discussed above, since the four possible solutions here are all  linearly independent, the same reasoning to the previous case applies and we conclude that this set of four observables also resolves the phase ambiguity.

Now, take the set $[(O^a_{1-}, O^a_{2-}) , (O^c_{1-}, O^c_{2-})]$. In this case, we obtain the following results: 
\begin{align}
    (\alpha_{13} - \alpha_{24}) & =  \ \ \, (\alpha_{12} - \alpha_{34})  \ , \nonumber \\
    (\alpha_{13} - \alpha_{24}) & =  - (\alpha_{12} - \alpha_{34})  \ , \nonumber \\
-  (\alpha_{13} - \alpha_{24}) & = \ \ \, (\alpha_{12} - \alpha_{34}) \ , \nonumber \\
-  (\alpha_{13} - \alpha_{24}) & = - (\alpha_{12} - \alpha_{34}) \ , 
\label{eq:2-28}
\end{align}
and we see that this set of observables cannot resolve the phase ambiguity, since there are degenerated (or linearly dependent) solutions (first and fourth solutions and second and third solutions).

Now, from Eqs.(\ref{eq:a1ma2m-phase},\ref{eq:a1pa2m-phase},\ref{eq:a1pa2p-phase},\ref{eq:a1ma2p-phase}), we note that the two relative phases, $\phi_{ij}$ and $\phi_{kl}$, involved in a given pair of observables from the same group, have the following properties ($m=a,b,c$):
\begin{widetext}
\begin{align}
  (O^m_{1\pm}, O^m_{2\mp}) = (O^m_{1+}, O^m_{2-})\ {\rm or}\ (O^m_{1-}, O^m_{2+})   & \longrightarrow \left\{ 
\begin{array}{llr}
\phi_{ij}  - \phi_{kl} \  \longrightarrow  \  \zeta{\rm -dependent}\ \ \, \ , \\    
\phi_{ij}  + \phi_{kl} \  \longrightarrow  \ \zeta{\rm -independent} \ ,    
\end{array} \right.  
\nonumber \\  
 (O^m_{1\pm}, O^m_{2\pm}) =  (O^m_{1+}, O^m_{2+})\ {\rm or}\ (O^m_{1-}, O^m_{2-}) &  \longrightarrow \left\{
\begin{array}{llr}
\phi_{ij}  - \phi_{kl} \   \longrightarrow \  \zeta{\rm -independent} \ , \\    
\phi_{ij}  + \phi_{kl} \  \longrightarrow \ \zeta{\rm -dependent} \ \ \, \ .    
\end{array} \right. \nonumber \\
\label{eq:2-29}
\end{align}
\end{widetext}

Then, from the pattern exhibited by the above three sets of observables worked out explicitly and with the help of Eq.(\ref{eq:2-29}), we can easily determine those sets of two pairs of observables for case (bb) that cannot resolve the phase ambiguity. They are the sets  which yield the phase relations in Eq.(\ref{eq:2-20}) being $\zeta$-independent. All the other sets do resolve the ambiguity. The results are displayed  in Table.\ref{tab:bb}.
\begin{turnpage}
\begin{table*}[t!]
\caption{\label{tab:bb} Sets of two pairs of double-spin observables for case (bb) mentioned in the text.  $\surd =$ do resolve. $X =$ do not resolve. Observables indicated outside the parentheses are the additional ones required in case the equal-relative-phase-magnitudes condition, as given by Eq.(\ref{eq:3-0}), is met for the pairs of observables (in parentheses) that do resolve the phase ambiguity otherwise.  }
\begingroup
\begin{tabular}{ccccccccc}
\hline\hline
  & $(O^b_{1+}, O^b_{2+}), O^b_{1-}\ $ & $(O^b_{1+}, O^b_{2-}), O^b_{1-}\ $ & $(O^b_{1-}, O^b_{2+}), O^b_{1+}\ $ & $(O^b_{1-}, O^b_{2-}), O^b_{1+}\ $ 
  & $(O^c_{1+}, O^c_{2+}), O^c_{1-}\ $ & $(O^c_{1+}, O^c_{2-}), O^c_{1-}\ $ & $(O^c_{1-}, O^c_{2+}), O^c{1+}\ $ & $(O^c_{1-}, O^c_{2-}), O^c{1+}$  \\
  & $(O_z, C_z), C_x$                & $(O_z, O_x), C_x$               & $(C_x, C_z), O_z$              & $(C_x, O_x), O_z$ 
  & $(L_x, L_z), T_z$                & $(L_x, T_x), T_z$                & $(T_z, L_z), L_x $               & $(T_z, T_x), L_x$  \\ \hline
$(O^a_{1+}, O^a_{2+}), O^a_{1-}$ & $\surd$   & $\surd$  & $\surd$  & $\surd$  &      X      & $\surd$  & $\surd$  &        X    \\
$(G, E), F $                       &    &   &   &   &   &   &   &      \\
$(O^a_{1+}, O^a_{2-}), O^a_{1-}$ & $\surd$   &      X       &       X      & $\surd$  & $\surd$  & $\surd$  & $\surd$  &  $\surd$  \\
$(G, H), F$                       &    &   &   &   &   &   &   &      \\
$(O^a_{1-}, O^a_{2+}), O^a_{1+}$ & $\surd$   &      X       &       X      & $\surd$  & $\surd$  & $\surd$  & $\surd$  &  $\surd$  \\
$(F, E), G$                       &    &   &   &   &   &   &   &      \\
$(O^a_{1-}, O^a_{2-}), O^a_{1+}$ & $\surd$   & $\surd$  & $\surd$  & $\surd$  &      X      & $\surd$  & $\surd$  &        X    \\
$(F, H), F$                       &    &   &   &   &   &   &   &      \\
  &    &   &   &   &   &   &   &      \\
$(O^c_{1+}, O^c_{2+}), O^c_{1-}$ & $\surd$   & $\surd$  & $\surd$  & $\surd$  &  &  &  &    \\
$(L_x, L_z), T_z$  &    &   &   &   &   &   &   &      \\
$(O^c_{1+}, O^c_{2-}), O^c_{1-}$ &     X         & $\surd$  & $\surd$  &      X       &  &  &   &     \\
 $(L_x, T_x), T_z$  &    &   &   &   &   &   &   &      \\
$(O^c_{1-}, O^c_{2+}), O^c_{1+}$ &     X        & $\surd$  & $\surd$  &       X      &   &   &   &    \\
$(T_z, L_z), L_x$   &    &   &   &   &   &   &   &      \\
$(O^c_{1-}, O^c_{2-}), O^c_{1+}$ & $\surd$   & $\surd$  & $\surd$  & $\surd$  &   &   &   &    \\
$(T_z, T_x), L_x$   &    &   &   &   &   &   &   &      \\
\hline\hline
\end{tabular}
\endgroup
\end{table*}
\end{turnpage}

\vskip 0.5cm
It should be noted, however, that there is a restriction to the fact that those sets of two pairs of observables can resolve the phase ambiguity. 
For example, for the set $[(O^a_{1+}, O^a_{2-}) , (O^c_{1-}, O^c_{2-})]$, from Eqs.(\ref{eq:2-26},\ref{eq:2-27}), it is clear that 
when $\alpha_{13} = - \alpha_{24}$ and/or $\alpha_{12} = \alpha_{34}$, no ambiguity can be resolved 
since the possible solutions become degenerated. The same is true for the set $[(O^a_{1+}, O^a_{2-}) , (O^c_{1+}, O^c_{2-})]$ when $\alpha_{13} = - \alpha_{24}$ and/or $\alpha_{12} = -\alpha_{34}$. 
It is easy to see that, had we considered the set  $[(O^a_{1+}, O^a_{2+}) , (O^c_{1+}, O^c_{2+})]$ instead, we would have found that when $\alpha_{13} =  \alpha_{24}$ and/or $\alpha_{12} =  \alpha_{34}$ no phase ambiguity can be resolved (cf.~Eqs.(\ref{eq:a1pa2p-phase},\ref{eq:2-20}a)).  
Thus, in these situations, we need to measure one or two more extra observables to be able to resolve the phase ambiguity. For example, for the set of two pairs of observables $[(O^a_{1+}, O^a_{2-}) , (O^c_{1-}, O^c_{2-})]$,  we require the extra observable $O^a_{1-}$ to resolve the ambiguity in the case $\alpha_{13} = - \alpha_{24}$ and, the extra observable $O^c_{1+}$  in the case $\alpha_{12} = \alpha_{34}$. 
If $\alpha_{13} = - \alpha_{24}$ and $\alpha_{12} = \alpha_{34}$, simultaneously, then, we require both extra observables 
$O^a_{1-}$ and $O^c_{1+}$.
Note that $O^a_{1+}$ differs by a sign of relative phase $\phi_{24}$ from $O^a_{1-}$.  This later feature is true for all the observables of the form $O^m_{1\nu}$. Thus, for the sets of two pairs of the form  
$[(O^a_{1\pm}, O^a_{2\nu}) , (O^c_{1\pm}, O^c_{2\nu'})]$, we need the extra observable $O^a_{1\mp}$ and/or $O^c_{1\mp}$ (here the $\pm$ signs are not independent) to completely resolve the phase ambiguity, depending on whether  $\alpha_{13} = \pm \alpha_{24}$ and/or $\alpha_{12} = \pm \alpha_{34}$. This means that we need a minimum of five or six chosen observables, instead of four, to resolve the phase ambiguity in these situations of equal magnitudes of the relative phases $\alpha_{ij}$'s.  It is straightforward to extended the above considerations to other sets of two pairs of observables involving groups $a$ and $b$, and groups $b$ and $c$.  The results are given in Table.~\ref{tab:bb}. 
Explicitly, the equal relative-phase magnitudes relations for the sets of two pairs of observables, in general, are
\begin{align}
| \alpha_{13} | & = |\alpha_{24} | \ \ {\rm and/or}\ \   | \alpha_{12} | = | \alpha_{34} | \qquad ( a \longleftrightarrow c) \ , \nonumber \\
| \alpha_{13} | & = |\alpha_{24} | \ \ {\rm and/or}\ \   | \alpha_{14} | = | \alpha_{23} | \qquad ( a \longleftrightarrow b) \ , \nonumber \\
| \alpha_{12} | & = |\alpha_{34} | \ \ {\rm and/or}\ \   | \alpha_{14} | = | \alpha_{23} | \qquad ( c \longleftrightarrow b) \ .  \nonumber \\
\label{eq:3-0}
\end{align}

Even with the additional observables as discussed above, the ambiguity still will not be resolved if $\alpha_{13}=\alpha_{24}=0$ and/or $\alpha_{12}=\alpha_{34}=0$. 
The only way to resolve the phase ambiguity in this case is to measure a set of eight chosen double-spin observables to determine both $\cos\phi_{ij}$ and $\sin\phi_{ij}$ for all four relative phases $\phi_{ij}$'s associated with the four basic photoproduction amplitudes.

\subsection{Case (ab)}
We now turn out attention to case (ab). In this case, it is straightforward to see that any pair of double-spin observables belonging to item (ii) that leads to
the corresponding phase relations as given by Eq.(\ref{eq:2-20}) being $\zeta$-dependent, resolves the phase ambiguity, irrespective of the pair of observables belonging to item (i). Otherwise the phase ambiguity cannot be resolved.  The results are displayed in Table.\ref{tab:ab}.
\begin{turnpage}
\begin{table*}[t!]
\caption{\label{tab:ab} Sets of two pairs of double-spin observables for case (ab) mentioned in the text.  $\surd =$ do resolve. $X =$ do not resolve. Observables indicated outside the parentheses are the additional ones required in case the equal-relative-phase-magnitudes condition, as given by Eq.(\ref{eq:3-0}), is met for the pairs of observables (in parenthese) that do resolve the phase ambiguity otherwise.  The additional observable required is either one of the observables indicated for each pair, except for those indicated with $**$, which require two additional observables. }
\begingroup
\begin{tabular}{ccccccc}
\hline\hline
   & $(O^a_{1+}, O^a_{1-}), O^a_{2\pm}\ $ & $(O^a_{2+}, O^a_{2-}), O^a_{1\pm}\ $ & $(O^b_{1+}, O^b_{1-}), O^b_{2\pm}\ $ & $(O^b_{2+}, O^b_{2-}), O^b_{1\pm}\ $ & $(O^c_{1+}, O^c_{1-}), O^c_{2\pm}\ $ & $(O^c_{2+}, O^c_{2-}), O^c_{1\pm}\ $ \\
   & $(G, F), E/H$                        & $(E, H), G/F $                         & $(O_z, C_x), C_z/O_x\ $                & $(C_z, O_x), O_z/C_x\ $               & $(L_x, T_z), L_z/T_x\ $               & $(L_z, T_x), L_x/T_z $                \\ \hline
$(O^a_{1+}, O^a_{2+}), O^a_{1-}$ &              &              & $\surd$\ \, **& $\surd$\ \, ** &     X      &     X      \\
$(G, E), F$                        & &  &  &  &  &  \\
$(O^a_{1+}, O^a_{2-}), O^a_{1-}$ &               &              &     X      &       X     & $\surd$ & $\surd$  \\
$(G, H), F$                        &  &  &  &  &  &  \\
$(O^a_{1-}, O^a_{2+}), O^a_{1+}$ &               &              &     X      &       X     & $\surd$ & $\surd$ \\
$(F, E), G$                          &  &  &  &  &  &  \\
$(O^a_{1-}, O^a_{2-}), O^a_{1+}$ &                &              & $\surd$\ \, ** & $\surd$\ \, ** &      X      &    X       \\
$(F, H), G$                         &   &  &  &  &  &  \\
  &  &  &  &  &  & \\
$(O^b_{1+}, O^b_{2+}), O^b_{1-}$ & $\surd$\ \, ** & $\surd$\ \, ** &             &               &       X    &     X       \\
$(O_z, C_z), C_x$                &  &  &  &  &  &  \\ 
$(O^b_{1+}, O^b_{2-}), O^b_{1-}$ &       X     &      X      &             &              & $\surd$\ \, **  & $\surd$\ \, ** \\
$(O_z, O_x), C_x$                &  &  &  &  &  &  \\
$(O^b_{1-}, O^b_{2+}), O^b_{1+}$ &       X     &      X      &             &              & $\surd$\ \, **  & $\surd$\ \, ** \\
$(C_x, C_z), O_z$               &  &  &  &  &  &   \\
$(O^b_{1-}, O^b_{2-}), O^b_{1+}$  & $\surd$\ \, ** &  $\surd$\ \, ** &            &               &       X     &     X      \\       
$(C_x, O_x), O_z$              &   &  &  &  &  &   \\
  &  &  &  &  &  & \\
$(O^c_{1+}, O^c_{2+}), O^c_{1-}$ &      X     &      X      & $\surd$\ \, ** & $\surd$\ \, ** &              &             \\
$(L_x, L_z), T_z$                 &  &  &  &  &  &  \\
$(O^c_{1+}, O^c_{2-}), O^c_{1-}$ & $\surd$ & $\surd$ &      X      &       X    &               &            \\
$(L_x, T_x), T_z$                 &  &  &  &  &  &  \\
$(O^c_{1-}, O^c_{2+}), O^c_{1+}$ &  $\surd$ & $\surd$ &     X     &       X     &               &            \\
$(T_z, L_z), L_x$               &  &  &  &  &  &  \\ 
$(O^c_{1-}, O^c_{2-}), O^c_{1+}$ &        X     &      X     & $\surd$\ \, ** & $\surd$\ \, ** &               &            \\
$(T_z, T_x), L_x$                &   &  &  &  &  &  \\
\hline\hline
\end{tabular}
\endgroup
\end{table*}
\end{turnpage}

Analogous to the previous case $(bb)$, here we have also the restriction of no equal relative-phase magnitudes, $|\alpha_{ij}|$'s, for the sets of two pairs of double-spin observables, as given in Table.~\ref{tab:ab}, to be able to resolve the phase ambiguity. This case  
involves the pairs of observables $(O^m_{n+}, O^m_{n-})\ (n=1,2)$, in addition to those encountered in case $(bb)$.   

In the case of $[(O^a_{1+}, O^a_{1-}) , (O^c_{1+} , O^c_{2+})]$, e.g., from Eqs.(\ref{eq:a1a1-phase},\ref{eq:2-20}a), the extra observable required to resolve the phase ambiguity is either $O^a_{2+}$ or $O^a_{2-}$ when $|\alpha_{13}| = |\alpha_{24}|$. 
Note that the relevant new pair of observables to help resolve the phase ambiguity here is either $(O^a_{1+}, O^a_{2+})$ or $(O^a_{1-}, O^a_{2-})$ (cf.~Eqs.(\ref{eq:a1pa2p-phase},\ref{eq:a1ma2m-phase})).  When $|\alpha_{12}| = |\alpha_{34}|$, the extra observable required is $O^c_{1-}$ as in case $(bb)$.

Now consider the set $[(O^a_{1+}, O^a_{1-}) , (O^b_{1+} , O^b_{2+})]$. In this case, from Eqs.(\ref{eq:a1a1-phase},\ref{eq:2-20}b), it requires both $O^a_{2+}$ and $O^a_{2-}$, in addition,  to resolve the phase ambiguity when $|\alpha_{13}| = |\alpha_{24}|$. And, as above, extra observable $O^b_{1-}$ when $|\alpha_{12}| = |\alpha_{34}|$.
 
For  the set $[(O^c_{1+}, O^c_{1-}) , (O^b_{1+} , O^b_{2+})]$, from Eqs.(\ref{eq:a1a1-phase},\ref{eq:2-20}c), it requires both $O^c_{2+}$ and $O^c_{2-}$ in addition, to resolve the phase ambiguity when $|\alpha_{13}| = |\alpha_{24}|$, and $O^b_{1-}$ in addition, when $|\alpha_{12}| = |\alpha_{34}|$. 

As for the two pairs of observables involving $(O^a_{2+}, O^a_{2-})$, from Eqs.(\ref{eq:a2a2-phase},\ref{eq:2-20}), we see that it always requires both $O^a_{1+}$ and $O^a_{1-}$ in addition, to resolve the phase ambiguity when $|\alpha_{13}| = |\alpha_{24}|$, irrespective of the other pair of observables from item (ii).  The latter, requires one extra observable when the corresponding relative phases have equal magnitudes.

We therefore see that in case $(ab)$, the minimum number of double-spin observables required to resolve the phase ambiguity - when the  magnitudes of the relative phases $\alpha_{ij}$ are equal - can be five, six or seven depending of the set of two pairs of observables that, otherwise, resolves the phase ambiguity. 
Based on the above considerations, the additional observables required to resolve the phase ambiguity are indicated in Table.~\ref{tab:ab}.

\section{Phase fixing for the $2+1+1$ case} \label{sec:211}

We start by considering two observables from a given group. For the sake of concreteness, consider the pair  $(O^a_{1+}, O^a_{1-}) = (-G, F)$.  This pair of observables was examined in the previous section with the phase ambiguity given 
in Eqs.(\ref{eq:2-2},\ref{eq:2-3}). Note that these two observables determine $\sin\phi_{13}$ and $\sin\phi_{24}$ (cf.~Eq.(\ref{eq:2-2})):
\begin{equation}
\sin\phi_{13}   = \frac{O^a_{1+} + O^a_{1-}}{2 B_{13}} \ , \qquad
\sin\phi_{24}  = \frac{O^a_{1+} - O^a_{1-}}{2 B_{24}}  \ .
\label{eq:1324}
\end{equation}
Appropriate combination of $\phi^\lambda_{24}$ and $\phi^{\lambda'}_{13}$ result in (cf. Eq.(\ref{eq:a1a1-phase}))
\begin{equation}
(O^a_{1+} , O^a_{1-}) : \  
 \left\{
\begin{array}{l}
\phi^+_{24} - \phi^+_{13}  = \ \ \, (\alpha_{24} - \alpha_{13}) \ ,  \\
\phi^+_{24} - \phi^-_{13}  =  \ \ \, (\alpha_{24} + \alpha_{13}) - \pi  \ ,  \\
\phi^-_{24} - \phi^+_{13}  = - (\alpha_{24} + \alpha_{13}) + \pi  \ ,  \\
\phi^-_{24} - \phi^-_{13}  = - (\alpha_{24} - \alpha_{13}) \ .  
\end{array} \right.  \\ \\
\label{eq:a1a1-phase1}
\end{equation}

\vskip 0.5cm
Now we consider two observables from the remaining two groups, $b={\cal BR}$ and $c={\cal TR}$.  For a given observable in one of these two groups, say $c={\cal TR}$, there will be four possible combinations of the pairs of observables one can form involving another observable from group $b={\cal BR}$ (cf Eqs(\ref{eq:comb1b},\ref{eq:comb1c})). For example, for the observable $O^c_{1+}$, we have the combinations $(O^b_{1-}, O^c_{1+})$, $(O^b_{1+}, O^c_{1+})$, $(O^b_{2-}, O^c_{1+})$, and $(O^b_{2+}, O^c_{1+})$.

\subsection{ $(O^b_{1\pm}, O^c_{1\pm})$ }

We start by considering the pair $(O^b_{1-}, O^c_{1+}) = (-C_x, -L_x)$. From Eqs.(\ref{eq:comb1b},\ref{eq:comb1c}),
\begin{align}
O^b_{1-} & = B_{14} \sin\phi_{14} - B_{23} \sin\phi_{23}  \ ,   \nonumber \\
O^c_{1+} & = B_{12} \sin\phi_{12} + B_{34} \sin\phi_{34}  \  .
\label{eq:CxLx}
\end{align}
Expressing $\phi_{14}$ and $\phi_{23}$ as
\begin{align}
\phi_{14} & = \phi_{24} + \phi_{12} \ , \nonumber \\
\phi_{23} & = \phi_{13} - \phi_{12} \ ,
\label{eq:1423}
\end{align}
we have 
\begin{equation}
O^b_{1-} = A_c \sin\phi_{12} + A_s \cos\phi_{12} \ ,  
\label{eq:AcAs0}
\end{equation}
with
\begin{align}
A_c & \equiv B_{14} \cos\phi_{24} + B_{23} \cos\phi_{13}   \ ,   \nonumber \\
A_s & \equiv B_{14} \sin\phi_{24} - B_{23} \sin\phi_{13}   \ .  
\label{eq:AcAs1}
\end{align}
Using $\cos\phi_{ij} = \pm \sqrt{1 - \sin^2\phi_{ij}}$\ , we solve Eq.(\ref{eq:AcAs0}) for $\sin\phi_{12}$ to obtain
\begin{equation}
\sin\phi_{12} = \frac{A_c O^b_{1-} \pm A_s \sqrt{D^2 - \left(O^b_{1-}\right){^2}}}{D^2} \ ,
\label{eq:12}
\end{equation}
with
\begin{equation}
D^2 \equiv A^2_c + A^2_s = B^2_{14} + B^2_{23}  + 2 B_{14} B_{23} \cos(\phi_{24} + \phi_{13}) \ .
\label{eq:D}
\end{equation}

We now note that while $A_s$  is uniquely determined (cf.~Eq.(\ref{eq:1324})), $A_c$ has a four-fold ambiguity because knowing only $\sin\phi_{ij}$ implies that $\cos\phi_{ij}$ is known up to a sign.  In particular, according to the notation of (\ref{eq:2-3}),
\begin{equation}
{\rm knowing}\ \sin\phi_{ij}^\lambda  \Longrightarrow \cos\phi^\lambda_{ij} = \lambda \cos\alpha_{ij} \ .
\label{eq:sclambda}
\end{equation}

Since $A_c$ depends on $\cos\phi^\lambda_{24}$ and $\cos\phi^{\lambda'}_{13}$ (cf.~Eq.(\ref{eq:AcAs1})), we introduce the notations $A^{\lambda \lambda'}_c$ and $D^{\lambda\lambda'\, 2}$, such that,
\begin{align}
A^{\lambda \lambda'}_c  &= B_{14} \cos\phi^\lambda_{24} + B_{23} \cos\phi^{\lambda'}_{13}  \ ,  \nonumber \\
D^{\lambda\lambda'\, 2} & = B^2_{14} + B^2_{23}  + 2 B_{14} B_{23} \cos(\phi^\lambda_{24} + \phi^{\lambda'}_{13}) \ .
\label{eq:AcLL'}
\end{align}
and, from Eq.(\ref{eq:12}), we see that $\phi_{12}$, in turn, depends on  $\lambda$ and $\lambda'$, i.e., 
\begin{equation}
\sin\phi^{\lambda \lambda'}_{12}(\eta) = \frac{A^{\lambda \lambda'}_c O^b_{1-} + \eta\, A_s \sqrt{D^{\lambda \lambda'\, 2}  - \left(O^b_{1-}\right){^2}}}{D^{\lambda \lambda'\, 2}} \ ,
\label{eq:12LL'}
\end{equation}
where $\eta$ takes the values $\pm 1$.

Due to Eq.(\ref{eq:sclambda}), it is clear that
\begin{align}
A^{+ +}_c & = - A^{- -}_c \qquad {\rm and} \qquad  \ \,  A^{+ -}_c = - A^{- +}_c \ , \nonumber \\
D^{+ +\, 2} & = D^{- -\, 2}  \qquad {\rm and} \qquad   D^{+ -\, 2} = D^{- +\, 2}  \ .
\label{eq:Ac}
\end{align}

Then, we have
\begin{align}
\sin\phi^{+ +}_{12}(\eta) & = \frac{A^{+ +}_c O^b_{1-} + \eta\, A_s \sqrt{D^{+ +\, 2} - \left(O^b_{1-}\right){^2}}}{D^{+ +\, 2}} \ ,  \nonumber \\
\sin\phi^{+ -}_{12}(\eta) & =  \frac{A^{+ -}_c O^b_{1-}  + \eta\, A_s \sqrt{D^{+ -\, 2} - \left(O^b_{1-}\right){^2}}}{D^{+ -\, 2}} \ ,  \nonumber \\
\sin\phi^{- +}_{12}(\eta) & = \frac{-A^{+ -}_c O^b_{1-} + \eta\, A_s \sqrt{D^{+ -\, 2} - \left(O^b_{1-}\right){^2}}}{D^{+ -\, 2}} \ ,  \nonumber \\
\sin\phi^{- -}_{12}(\eta) & = \frac{- A^{+ +}_c O^b_{1-} + \eta\, A_s \sqrt{D^{+ +\, 2} - \left(O^b_{1-}\right){^2}}}{D^{+ +\, 2}} \ . 
\label{eq:12LL'a}
\end{align}

From the above results, we see that there are, in general, eight possible $\sin\phi^{\lambda \lambda'}_{12}(\eta)$'s (recall that $\lambda, \lambda'$ and $\eta$ take two possible values each), and each of them leads to a two-fold ambiguity 
\begin{equation}
\phi^{\lambda\lambda'}_{12}(\eta) = \left\{
\begin{array}{l}
\alpha^{\lambda\lambda'}_{12}(\eta)  \ \ \ \ \ \ \ ,   \\
\pi - \alpha^{\lambda\lambda'}_{12}(\eta)  \ \ .
\end{array} \right. 
\label{eq:alpha12LL'}
\end{equation}
An inspection of Eq.(\ref{eq:12LL'a}) reveals that 
\begin{align}
\sin\phi^{+ +}_{12}(\pm) & = - \sin\phi^{- -}_{12}(\mp)  \ , \nonumber \\
\sin\phi^{+ -}_{12}(\pm) & = - \sin\phi^{- +}_{12}(\mp)  \ ,
\label{eq:2sn12-prop}
\end{align}
and, consequently,
\begin{equation}
\alpha^{+ +}_{12}(\pm) = - \alpha^{- -}_{12}(\mp)  \quad {\rm and} \quad \alpha^{+ -}_{12}(\pm) = - \alpha^{- +}_{12}(\mp)  \ .
\label{eq:alpha12-prop}
\end{equation}

Note that since all $\sin\phi^{\lambda \lambda'}_{12}(\eta)$'s are distinct from each other, so are $\alpha^{\lambda \lambda'}_{12}(\eta)$'s.

\vskip 0.5cm
Now, taking the equation for $O^c_{1+}$ in (\ref{eq:CxLx}) and solving for $\sin\phi_{34}$, yields
\begin{equation}
\sin\phi^{\lambda\lambda'}_{34}(\eta) = \frac{O^c_{1+} - B_{12} \sin\phi^{\lambda\lambda'}_{12}(\eta)}{B_{34}} \ ,
\label{eq:34}
\end{equation}
where we have displayed all the indices of the relative phases $\phi_{12}$ and $\phi_{34}$ explicitly. The above result leads to the two-fold ambiguity
\begin{equation}
\phi^{\lambda \lambda'}_{34}(\eta) = \left\{
\begin{array}{l}
\alpha^{\lambda \lambda'}_{34}(\eta)  \ ,   \\
\pi - \alpha^{\lambda \lambda'}_{34}(\eta)  \ ,
\end{array} \right.  
\label{eq:alpha34}
\end{equation}
with all eight $\alpha^{\lambda \lambda'}_{34}(\eta)$ being distinct from each other to the extent that $\sin\phi^{\lambda\lambda'}_{12}(\eta)$'s are. However, $\alpha^{\lambda\lambda'}_{34}(\eta)$ lacks the symmetry exhibited by $\alpha^{\lambda\lambda'}_{12}(\eta)$ in Eq.(\ref{eq:alpha12-prop}), i.e., $\alpha^{\lambda\lambda'}_{34}(\eta)$'s are not related to each other in general.

\vskip 0.5cm
Appropriate combinations of the relative phases  $\phi^{\lambda\lambda'}_{34}(\eta)$ and $\phi^{\lambda\lambda'}_{12}(\eta)$ involved in each pair contain, in general, a four-fold ambiguity of the form given by
\begin{equation}
\phi^{\lambda\lambda'}_{34}(\eta) - \phi^{\lambda\lambda'}_{12}(\eta)  =  \left\{
\begin{array}{l}
\ \ \, \left(\alpha^{\lambda \lambda'}_{34}(\eta) - \alpha^{\lambda \lambda'}_{12}(\eta) \right) \ ,  \\
 \ \  \, \left( \alpha^{\lambda \lambda'}_{34}(\eta) + \alpha^{\lambda \lambda'}_{12}(\eta) \right) - \pi  \ , \\
 - \left(\alpha^{\lambda \lambda'}_{34}(\eta) + \alpha^{\lambda \lambda'}_{12}(\eta) \right) + \pi  \ , \\
  - \left(\alpha^{\lambda \lambda'}_{34}(\eta) - \alpha^{\lambda \lambda'}_{12}(\eta) \right)  \ ,
\end{array} \right. 
\label{eq:blcl'-phasem}
\end{equation}
for a given set of $\{\lambda, \lambda', \eta\}$ (note that  $\lambda, \lambda'$ and $\eta$ take two possible values each).

\vskip 0.5cm
 At this stage, in analogy to what we have done in the \textbf{(2+2) case} in the previous section, we invoke the consistency relation (\ref{eq:2-20a}) reexpressed as (cf.~Eq.(\ref{eq:2-20}a))
\begin{equation}
\phi^\lambda_{24} - \phi^{\lambda'}_{13} =  \phi^{\lambda\lambda'}_{34}(\eta) - \phi^{\lambda \lambda'}_{12}(\eta) \ .
\label{eq:2413-3412}
\end{equation}
Inserting Eq.(\ref{eq:blcl'-phasem}) into the above equation, we arrive at the possible solutions 
\begin{equation}
\phi^{\lambda}_{24} - \phi^{\lambda'}_{13}  =  \left\{
\begin{array}{l}
 \ \ \, \left(\alpha^{\lambda \lambda'}_{34}(\eta) - \alpha^{\lambda \lambda'}_{12}(\eta) \right) \ ,  \\
\ \  \, \left( \alpha^{\lambda \lambda'}_{34}(\eta) + \alpha^{\lambda \lambda'}_{12}(\eta) \right) - \pi  \ , \\
 - \left(\alpha^{\lambda \lambda'}_{34}(\eta) + \alpha^{\lambda \lambda'}_{12}(\eta) \right) + \pi  \ , \\
- \left(\alpha^{\lambda \lambda'}_{34}(\eta) - \alpha^{\lambda \lambda'}_{12}(\eta) \right)  \ , 
\end{array} \right. 
\label{eq:solutions}
\end{equation}
for a given set of $\{\lambda, \lambda', \eta\}$. The left-hand-side of the above equation is given by Eq.(\ref{eq:a1a1-phase1}).
Since  $\lambda, \lambda'$ and $\eta$ take two possible values each,  we have  $2 \times 2 = 4$  distinct combinations on the left-hand-side of the above equation (cf.~Eq.(\ref{eq:a1a1-phase1}))
and, on the right-hand-side, we have $4 \times 2 = 8$ distinct combinations. This ends up with a total of $4 \times 8 = 32$ possible  solutions. It happens that these 32 solutions are all linearly independent, i.e., there are no degenerated solutions among them. This follows from the fact that all $\sin\phi^{\lambda\lambda'}_{34}(\eta)$'s  - and, in turn, all $\alpha^{\lambda\lambda'}_{34}(\eta)$'s  - are distinct from each other as pointed out previously (see below Eq.(\ref{eq:alpha34})). Thus, once the unique values of $\alpha_{13}$, $\alpha_{24}$ and the associated $\alpha^{\lambda\lambda'}_{12}(\eta)$  and $\alpha^{\lambda\lambda'}_{34}(\eta)$ are provided by the measurements of the observables $[(O^a_{1+}, O^a_{1+}), (O^b_{1-}, O^c_{1+})]$, there will be only one solution satisfying the consistency relation  (\ref{eq:2413-3412}). Therefore, we conclude that this set of observables will resolve the phase ambiguity.

\vskip 0.3cm
It is clear that the preceding results for the pair of observables $(O^b_{1-}, O^c_{1+})$, actually holds for any of the pairs $(O^b_{1\pm}, O^c_{1\pm})$, with the signs $\pm$ being independent,  
since the only difference is the sign change of $B_{23}$ and/or $B_{34}$ according to the particular combination of the observables in the pair considered. These sign changes do not affect any of the properties exhibited by the phases $\alpha^{\lambda\lambda'}_{12}(\eta)$ and  $\alpha^{\lambda\lambda'}_{34}(\eta)$. Thus, any one of the pairs of observables $(O^b_{1\pm} , O^c_{1\pm})$, together with the pair $(O^a_{1+} , O^a_{1-})$, can resolve the phase ambiguity of the transversity amplitude.

\subsection{$(O^b_{2 \pm} , O^c_{1\pm})$}  

We now consider  the pair $(O^b_{2-}, O^c_{1+}) = (-O_x, -L_x)$,
\begin{align}
O^b_{2-} & = B_{14} \cos\phi_{14} - B_{23} \cos\phi_{23}  \ ,   \nonumber \\
O^c_{1+} & = B_{12} \sin\phi_{12} + B_{34} \sin\phi_{34}  \  .
\label{eq:OxLx}
\end{align}
In this case, inserting Eq.(\ref{eq:1423}) into the expression for $O^b_{2-}$ in the above equation, yields
\begin{equation}
O^b_{2-} = A_c \cos\phi_{12} - A_s \sin\phi_{12} \ ,  
\label{eq:AcAs2}
\end{equation}
with
\begin{align}
A_c & \equiv B_{14} \cos\phi_{24} - B_{23} \cos\phi_{13}   \ ,   \nonumber \\
A_s & \equiv B_{14} \sin\phi_{24} + B_{23} \sin\phi_{13}   \ .  
\label{eq:AcAs3}
\end{align}

Solving Eq.(\ref{eq:AcAs2}) for $\sin\phi_{12}$, we have
\begin{equation}
\sin\phi_{12} = \frac{-O^b_{2-}\, A_s \pm A_c \sqrt{ D^2 - \left(O^b_{2-}\right){^2} }}{D^2} \ ,
\label{eq:25}
\end{equation}
where
\begin{equation}
D^2 \equiv A_c^2 + A_s^2  =  B^2_{14} + B^2_{23} - 2 B_{14} B_{23} \cos(\phi_{24} + \phi_{13}) \ .
\label{eq:25a}
\end{equation}

Using the same notation introduced in Eq.(\ref{eq:AcLL'}), we write Eq.(\ref{eq:25}) as 
\begin{equation}
\sin\phi^{\lambda\lambda'}_{12}(\eta) = \frac{-O^b_{2-}\, A_s + \eta\, A^{\lambda\lambda'}_c \sqrt{ D^{\lambda\lambda'\, 2} - \left(O^b_{2-}\right){^2} }}{D^{\lambda\lambda'\, 2} } \ .
\label{eq:27}
\end{equation}
Noticing that both $A^{\lambda\lambda'}_c$  and $D^{\lambda\lambda'\, 2}$ here have the same symmetry as in Eq.(\ref{eq:Ac}), we can verify in this case that
\begin{align}
\sin\phi^{+ +}_{12}(\pm) & = \sin\phi^{- -}_{12}(\mp)  \ , \nonumber \\
\sin\phi^{+ -}_{12}(\pm) & = \sin\phi^{- +}_{12}(\mp)  \ ,
\label{eq:2sn12-propa}
\end{align}
and, consequently,
\begin{align}
\alpha^{+ +}_{12}(\pm) & = \alpha^{- -}_{12}(\mp)  \quad {\rm and} \quad \alpha^{+ -}_{12}(\pm) = \alpha^{- +}_{12}(\mp)  \ . \nonumber \\
\label{eq:alpha12-propa}
\end{align}
Also, note that for a given set of $\{\lambda, \lambda', \eta\}$, Eq.(\ref{eq:27}) leads to a two-fold phase ambiguity  as given by Eq.(\ref{eq:alpha12LL'}).

Solving now the equation for $O^c_{1+}$ in (\ref{eq:OxLx}) for $\sin\phi_{34}$, we have
\begin{equation}
\sin\phi^{\lambda\lambda'}_{34}(\eta) = \frac{O^c_{1+} - B_{12} \sin\phi^{\lambda\lambda'}_{12}(\eta)}{B_{34}} \ ,
\label{eq:34a}
\end{equation}
leading to a two-fold phase ambiguity as given by Eq.(\ref{eq:alpha34}).
Here we note that, unlike in the case of the pair of observables $(O^b_{1-}, O^c_{1+})$, where $\sin\phi^{\lambda\lambda'}_{34}(\eta)$ has no symmetry, this quantity given by Eq.(\ref{eq:34a}) above exhibits the following symmetry:
\begin{align}
\sin\phi^{+ +}_{34}(\pm) & = \sin\phi^{- -}_{34}(\mp) \ , \nonumber \\
\sin\phi^{+ -}_{34}(\pm) & = \sin\phi^{- +}_{34}(\mp) \ ,
\label{eq:sn34-sym}
\end{align}
where Eq.(\ref{eq:2sn12-propa}) has been used. Consequently,
\begin{equation}
\alpha^{+ +}_{34}(\pm) = \alpha^{- -}_{34}(\mp) \quad {\rm and} \quad   \alpha^{+ -}_{34}(\pm) = \alpha^{- +}_{34}(\mp) \ .
\label{eq:alpha34-sym}
\end{equation}

The relative phases $\alpha^{\lambda\lambda'}_{12}(\eta)$ and $\alpha^{\lambda\lambda'}_{34}(\eta)$ derived here, with the symmetry properties given by Eqs.(\ref{eq:alpha12-propa},\ref{eq:alpha34-sym}), should obey Eq.(\ref{eq:solutions}).  It happens that the set of pairs $[(O^b_{2-}, O^c_{1+}) , (O^a_{1+}, O^a_{1-})]$ cannot resolve the phase ambiguity. To see this,
it suffices to consider the following two particular solutions from Eq.(\ref{eq:solutions}), 
\begin{widetext}
\begin{align}
\phi^+_{24} - \phi^+_{13} & =  \alpha^{+ +}_{34}(+) -  \alpha^{+ +}_{12}(+) \quad \Longrightarrow \qquad    \left(\alpha_{24} - \alpha_{13} \right)  =  \quad \left(\alpha^{+ +}_{34}(+) -  \alpha^{+ +}_{12}(+) \right) \ , \nonumber \\
\phi^-_{24} - \phi^-_{13} & =  \alpha^{- -}_{34}(-) -  \alpha^{- -}_{12}(-) \quad \Longrightarrow \quad   -\left(\alpha_{24} - \alpha_{13} \right)  =   -\left(\alpha^{+ +}_{34}(+) - \alpha^{+ +}_{12}(+) \right) \ ,  
\label{eq:1234-check}
\end{align}
\end{widetext}
where we have made use of Eqs.(\ref{eq:a1a1-phase1},\ref{eq:alpha12-propa},\ref{eq:alpha34-sym}).  This shows that these solutions are linearly dependent (degenerated) and, consequently, the set of observables in consideration cannot resolve the phase ambiguity. Degeneracy of the solutions involving $\alpha^{+ -}_{ij}(\pm)$ and $\alpha^{- +}_{ij}(\mp)$ also occurs.

\vskip 0.3cm
The above consideration shows that any of the pairs of observables  $(O^b_{2\pm}, O^c_{1\pm})$, together with the pair $(O^a_{1+} , O^a_{1-})$, cannot resolve the phase ambiguity of the transversity amplitude.

\subsection{$(O^b_{2\pm} , O^c_{2\pm})$}

For $(O^b_{2-} , O^c_{2+})=(-O_x, - L_z)$, 
\begin{align}
O^b_{2-} & = B_{14} \cos\phi_{14} - B_{23} \cos\phi_{23}  \ ,   \nonumber \\
O^c_{2+} & = B_{12} \cos\phi_{12} + B_{34} \cos\phi_{34}  \  ,
\label{eq:OxLz}
\end{align}
proceeding analogously to the case of  $(O^b_{1-}, O^c_{1+})$, we have 
\begin{equation}
\cos\phi^{\lambda \lambda'}_{12}(\eta) = \frac{A^{\lambda \lambda'}_c O^b_{2-} + \eta\, A_s \sqrt{D^{\lambda \lambda'\, 2}  - \left(O^b_{2-}\right){^2}}}{D^{\lambda \lambda'\, 2}} \ ,
\label{eq:cs12LL'}
\end{equation}
where
\begin{align}
A_s  &= B_{14} \sin\phi_{24} + B_{23} \sin\phi_{13}  \ ,  \nonumber \\
A^{\lambda \lambda'}_c  &= B_{14} \cos\phi^\lambda_{24} - B_{23} \cos\phi^{\lambda'}_{13}  \ ,  \nonumber \\
D^{\lambda\lambda'\, 2} & = B^2_{14} + B^2_{23}  - 2 B_{14} B_{23} \cos(\phi^\lambda_{24} + \phi^{\lambda'}_{13}) \ .
\label{eq:csAcLL'}
\end{align}
It is clear that $\cos^{\lambda\lambda'}_{12}(\eta)$ above exhibits the symmetry
\begin{align}
\cos\phi^{+ +}_{12}(\pm) & = - \cos\phi^{- -}_{12}(\mp)  \ , \nonumber \\
\cos\phi^{+ -}_{12}(\pm) & = - \cos\phi^{- +}_{12}(\mp)  \ ,
\label{eq:2cs12-prop}
\end{align}
and, consequently,
\begin{equation}
\alpha^{+ +}_{12}(\pm) = \pi + \alpha^{- -}_{12}(\mp)   \quad {\rm and} \quad \alpha^{+ -}_{12}(\pm) = \pi + \alpha^{- +}_{12}(\mp)   \ . 
\label{eq:csalpha12-prop}
\end{equation}
\vskip 0.3cm
Now, solving the equation for $O^c_{2+}$ in (\ref{eq:OxLz}) for $\cos\phi_{23}$, yields
\begin{align}
\cos\phi^{\lambda\lambda'}_{34}(\eta) & = \frac{O^c_{2+} - B_{12}\cos\phi^{\lambda\lambda'}_{12}(\eta)}{B_{34}} \ , 
\label{eq:cs34LL'} 
\end{align}
which reveals that all eight possible values of it are distinct. Consequently, all $\alpha^{\lambda\lambda'}_{34}(\eta)$'s are distinct, resulting in linear independence of all  possible solutions from the consistency relation (\ref{eq:2-20}a).   Then, it follows that, any pair of observables of the form $(O^b_{2\pm} , O^c_{2\pm})$,  together with the pair $(O^a_{1+} , O^a_{1-})$ can resolve the phase ambiguity.

\vskip 0.5cm
Summarizing the results obtained in this section so far, we have
\begin{widetext}
\begin{align}
(O^a_{1+}, O^a_{1-})\ {\rm and}\ (O^b_{n\pm}, O^c_{n \pm})\ (n=1,2)  & \to  {\rm do\ resolve\ the\ ambiguity} \  , \nonumber \\ 
(O^a_{1+},O^a_{1-}) \ {\rm and}\   (O^b_{2 \pm}, O^c_{1 \pm})  & \to  {\rm do\ not\ resolve\ the\ ambiguity} \  .
\label{eq:snsn-result}
\end{align}
\end{widetext}
In the above relations, the $\pm$ signs are independent.

\subsection{ $(O^a_{2+} , O^a_{2-})$}

We now turn our attention to the case of the pair of observables from group $a$ being $(O^a_{2+}, O^a_{2-}) = (E, H)$,
\begin{align}
O^a_{2+} & = B_{13} \cos\phi_{13} + B_{24} \cos\phi_{24}  \ ,   \nonumber \\
O^a_{2-} & = B_{13} \cos\phi_{13} - B_{24} \cos\phi_{24}  \  .
\label{eq:EH}
\end{align}
The difference from the previous case of $(O^a_{1+}, O^a_{1-})$ is that $(O^a_{2+}, O^a_{2 -})$ determines $\cos\phi_{24}$ and $\cos\phi_{13}$ uniquely, instead of $\sin\phi_{24}$ and $\sin\phi_{13}$.  This implies that, for the pair $(O^b_{ 1-} , O^c_{1+})$, 
the quantity $A_c$ defined in Eq.(\ref{eq:AcAs1}) becomes uniquely determined, while  $A_s$ will have a four-fold ambiguity and the quantity $D^2$ in Eq.(\ref{eq:D}) depends on $(\lambda\lambda')$, but remains unchanged otherwise, viz., 
\begin{align}
A_c  &= B_{14} \cos\phi_{24} + B_{23} \cos\phi_{13}  \ ,  \nonumber \\
A^{\lambda \lambda'}_s  &= B_{14} \sin\phi^\lambda_{24} - B_{23} \sin\phi^{\lambda'}_{13}  \ ,  \nonumber \\
D^{\lambda\lambda'\, 2} & = B^2_{14} + B^2_{23}  + 2 B_{14} B_{23} \cos(\phi^\lambda_{24} + \phi^{\lambda'}_{13}) \ .
\label{eq:AsAcDLL'}
\end{align}
Then,  Eq.(\ref{eq:12LL'})  changes to
\begin{equation}
\sin\phi^{\lambda \lambda'}_{12}(\eta) = \frac{A_c O^b_{1-} + \eta\, A^{\lambda\lambda'}_s \sqrt{D^{\lambda \lambda'\, 2}  - \left(O^b_{1-}\right){^2}}}{D^{\lambda \lambda'\, 2}} \ .
\label{eq:12LL'a2}
\end{equation}

Analogously, for the pair $(O^b_{ 2-} , O^c_{1+})$, Eq. (\ref{eq:27}) changes to
\begin{equation}
\sin\phi^{\lambda\lambda'}_{12}(\eta) = \frac{-O^b_{2-}\, A^{\lambda\lambda'}_s + \eta\, A_c \sqrt{ D^{\lambda\lambda'\, 2} - \left(O^b_{2-}\right){^2} }}{D^{\lambda\lambda'\, 2} } \ .
\label{eq:27a2}
\end{equation}
In the above equation $A_c, A_c^{\lambda\lambda'}$ and $D^{\lambda\lambda'\, 2}$ are given by Eq.(\ref{eq:AsAcDLL'}) except for the change in the sign of $B_{23}$.

It, then, follows that the symmetry properties of $\sin\phi^{\lambda\lambda'}_{12}(\eta)$ given in the above two equations have interchanged from the corresponding quantities in the case of $(O^a_{1+}, O^a_{1-})$. This, in turn, interchanges the property of $\alpha^{\lambda\lambda'}_{34}(\eta)$. We can now see
that the role of $(O^b_{1\pm}, O^c_{1 \pm})$ and $(O^b_{2 \pm}, O^c_{1 \pm})$ interchanges in 
Eq.(\ref{eq:snsn-result}), i.e.,
\begin{widetext}
\begin{align}
(O^a_{2+}, O^a_{2 -})\ {\rm and}\ (O^b_{n\pm}, O^c_{n \pm})\ (n=1,2) & \to  {\rm do\ not\ resolve\ the\ ambiguity} \  , \nonumber \\ 
(O^a_{2+},O^a_{2-}) \ {\rm and}\   (O^b_{2 \pm}, O^c_{1 \pm})  & \to  {\rm do\  resolve\ the\ ambiguity} \  .
\label{eq:cscs-result}
\end{align}
\end{widetext}

\subsection{$(O^a_{1\pm} , O^a_{2\pm})$}

In the case of $(O^a_{1\pm} , O^a_{2 \mp})$ (here the signs $\pm$ are not independent), we note that 
 $\phi^\lambda_{24} -\phi^{\lambda'}_{13}$ is $\zeta$-dependent (cf. Eqs.(\ref{eq:a1pa2m-phase},\ref{eq:a1ma2p-phase})). Therefore, in this case, the phase ambiguity will be resolved because
the possible solutions in Eq.(\ref{eq:solutions}) will all be linearly independent. 
 For the case of $(O^a_{1 \pm} , O^a_{2 \pm})$ (not independent $\pm$ signs), however, $\phi^\lambda_{24} -\phi^{\lambda'}_{13}$ is $\zeta$-independent (cf. Eqs.(\ref{eq:a1ma2m-phase},\ref{eq:a1pa2p-phase})) and the  above argument valid for $(O^a_{1\pm} , O^a_{2 \mp})$ does not apply. However, it happens that the relative phases $\phi_{24}$ and $\phi_{13}$ in the $(O^a_{1\pm} , O^a_{2\pm})$ case are given by 
(cf. Eqs(\ref{eq:2-18a},\ref{eq:2-18})) 
\begin{align}
& \left\{
\begin{array}{ll} 
\phi_{13} = & - \zeta + \alpha_{13} \ ,  \\
\phi_{24} = & - \zeta + \alpha_{24} - \delta_{+} \pi \ ,
\end{array} \right. \\
\quad {\rm or} \quad \nonumber \\
& \left\{
\begin{array}{ll} 
\phi_{13} = &  - \zeta - \alpha_{13} + \pi \ , \\
 \phi_{24} = & - \zeta  - \alpha_{24} + \delta_- \pi     \ ,
 \end{array} \right.
\label{eq:2-18pp}
\end{align}
with two-fold ambiguity.  $\delta_+=1$ and $\delta_-=0$ for $(O^a_{1+} , O^a_{2+})$ and $\delta_+=0$ and $\delta_-=1$ for $(O^a_{1-} , O^a_{2-})$. 
It is then easy to see that all $\cos\phi_{ij}$ ($ij=24, 34$) are distinct from each other. The same is true for $\sin\phi_{ij}$.  This implies that the quantities $A_c$ and $A_s$ entering into Eqs.(\ref{eq:12LL'},\ref{eq:27}) have  all distinct values, in general, as can be seen from their definitions in Eqs.(\ref{eq:AcAs1},\ref{eq:AcAs3}) for the case $(O^b_{1-} , O^c_{1+})$ and $(O^b_{2-} , O^c_{1+})$, respectively. Hence, all the phases $\alpha^{\lambda\lambda'}_{12}(\eta)$ and  $\alpha^{\lambda\lambda'}_{34}(\eta)$ entering into Eq.(\ref{eq:solutions}) assume distinct values in general, resulting in linearly independent possible solutions. Consequently, the phase ambiguity can be resolved with the pairs $(O^a_{1 \pm} , O^a_{2 \pm})$ as well.  

We conclude that any pair of the form  $(O^a_{1\pm} , O^a_{2\pm})$,  together with any pair of the form $(O^b_{1\pm} , O^c_{1\pm})$ or  $(O^b_{2\pm} , O^c_{1\pm})$, will resolve the phase ambiguity. Here all the signs $\pm$ are independent.

\vskip 1cm
This completes the analysis of all possible \textbf{($2 + 1 + 1$) cases}. Collecting the results for all the possibilities,   
the following sets of four observables will resolve the phase ambiguity in the \textbf{($2 + 1 + 1$) case} :
\begin{itemize}

\item[i)]
 $(O^a_{1 +}\,, O^a_{1 -})$ and $\left[ (O^b_{1\pm} , O^c_{1\pm}) \ {\rm or}\  (O^b_{2\pm} , O^c_{2\pm}) \right]$.

\item[ii)]
 $(O^a_{1 \pm}\,, O^a_{2 \pm})$ and  $\left[ (O^b_{1\pm} , O^c_{1\pm}) \ {\rm or}\  (O^b_{2\pm} , O^c_{2\pm}) \ {\rm or}\right. \\  \left. (O^b_{2\pm} , O^c_{1\pm}) \right]$.

\item[iii)]
 $(O^a_{2 +}\,, O^a_{2 -})$ and  $(O^b_{2\pm} , O^c_{1\pm})$.

\end{itemize}
with any permutation of $a, b, c$. Here, the $\pm$ signs are all independent. The results are displayed in Table.~\ref{tab:211} for the \textbf{case ($2 (a) + 1 (b) + 1 (c)$)}. Other combinations can be obtained by an appropriate permutation of $a, b, c$.
\begin{table*}[t!]
\caption{\label{tab:211} Sets of two pairs of double-spin observables for \textbf{case ($2 (a) + 1 (b) + 1 (c)$)}. Other combinations can be obtained by appropriate permutations of the indices $a, b, c$.  $\surd =$ do resolve. $X =$ do not resolve. Observables indicated outside the parentheses are the additional ones required in case the equal-relative-phase-magnitudes condition, as given by Eq.(\ref{eq:3-0}), is met for the pairs of observables (in parentheses) that do resolve the phase ambiguity otherwise. 
 The additional observable required is either one of the observables indicated for each pair, except for those marked with  $**$, which require any two additional observables from those indicated. }
\begin{tabular}{ccccccc}
\hline\hline
  & $(O^a_{1+}, O^a_{1-}), O^a_{2\pm}\ $ & $(O^a_{1+}, O^a_{2+}), O^a_{1-}\ $ & $(O^a_{1+}, O^a_{2-}), O^a_{1-}\ $ & $(O^a_{1-}, O^a_{2+}), O^a_{1+}\ $ 
  & $(O^a_{1-}, O^c_{2-}), O^a_{1+}\ $ & $(O^a_{2+} , O^a_{2-}), O^a_{1\pm}$  \\ 
  & $(G, F), E/H$                & $(G, E), F$               & $(G, H), F$              & $(F, E), G$ 
  & $(F, H), G$                & $(E, H), G/F$  \\ \hline
$O^b_{1-}/O^c_{1-}, (O^b_{1+}, O^c_{1+}) $ & $\surd$  &  $\surd$  &  $\surd$    & $\surd$  &    $\surd$     &  X      \\
$C_x/T_z,  (O_z, L_x) $                       &    &   &   &   &   &         \\
 $O^b_{1-}/O^c_{1+}, (O^b_{1+}, O^c_{1-}) $ & $\surd$   &    $\surd$    &   $\surd$   & $\surd$  & $\surd$  & X    \\
$C_x/L_x, (O_z, T_z) $                       &    &   &   &   &   &         \\
$O^b_{1-}, (O^b_{1+}, O^c_{2+}) $ & X      &      $\surd$   &    $\surd$      & $\surd$  & $\surd$  & $\surd$\ \, **    \\
$C_x, (O_z, L_z) $                       &    &   &   &   &   &        \\
$O^b_{1-}, (O^b_{1+}, O^c_{2-}) $ & X       &     $\surd$  &        $\surd$  &   $\surd$  &  $\surd$   &  $\surd$\ \, **     \\
$C_x, (O_z, T_x) $                       &    &   &   &   &   &         \\
  &    &   &   &   &   &         \\
$O^b_{1+}/O^c_{1-}, (O^b_{1-}, O^c_{1+}) $ & $\surd$   & $\surd$  & $\surd$  & $\surd$  & $\surd$  &  X    \\
$O_z/T_z, (C_x, L_x) $  &    &   &   &   &   &      \\
$O^b_{1+}/O^c_{1+}, (O^b_{1-}, O^c_{1-}) $ & $\surd$   & $\surd$  & $\surd$  & $\surd$  & $\surd$ &  X   \\
$O_z/L_x, (C_x, T_z) $  &    &   &   &   &   &      \\
$O^b_{1+}, (O^b_{1-}, O^c_{2+}) $ & X  & $\surd$  & $\surd$  & $\surd$  &  $\surd$ &  $\surd$ \ \, **   \\
$O_z, (C_x, L_z) $  &    &   &   &   &   &      \\
$O^b_{1+}, (O^b_{1-}, O^c_{2-}) $ &     X         & $\surd$  & $\surd$  &   $\surd$     & $\surd$  &  $\surd$\ \, **     \\
 $O_z, (C_x, T_x) $  &    &   &   &   &   &       \\
   &    &   &   &   &   &         \\
$O^c_{1-}, (O^b_{2+}, O^c_{1+}) $ & X   & $\surd$  & $\surd$  & $\surd$  & $\surd$  &  $\surd$ \ \, **   \\
$T_z, (C_z, L_x) $  &    &   &   &   &   &      \\
$O^c_{1+} (O^b_{2+}, O^c_{1-}) $ & X   & $\surd$  & $\surd$  & $\surd$  & $\surd$ &  $\surd$\ \, **  \\
$L_x, (C_z, T_z) $  &    &   &   &   &   &      \\
$O^b_{1\pm}/O^c_{1\pm}, (O^b_{2+}, O^c_{2+}) $ & **\ \, $\surd$  & **\ \, $\surd$  & $\surd$  & $\surd$  &  $\surd$ &  X    \\
$(C_z, L_z)$  &    &   &   &   &   &      \\
$O^b_{1\pm}/O^c_{1\pm}, (O^b_{2+}, O^c_{2-}) $ &  **\ \,  $\surd$    & **\ \, $\surd$  & $\surd$  &   $\surd$     & $\surd$  &  X    \\
 $(C_z, T_x)$  &    &   &   &   &   &       \\
  &    &   &   &   &   &         \\
$O^c_{1-}, (O^b_{2-}, O^c_{1+}) $ & X   & $\surd$  & $\surd$  & $\surd$  & $\surd$  &  $\surd$ \ \, **   \\
$T_z, (O_x, L_x) $  &    &   &   &   &   &      \\
$O^c_{1+}, (O^b_{2-}, O^c_{1-}) $ & X   & $\surd$  & $\surd$  & $\surd$  & $\surd$ &  $\surd$\ \, **   \\
$L_x, (O_x, T_z) $  &    &   &   &   &   &      \\
$O^b_{1\pm}/O^c_{1\pm}, (O^b_{2-}, O^c_{2+}) $ & **\ \, $\surd$  & **\ \, $\surd$  & $\surd$  & $\surd$  &  $\surd$ &  X    \\
$(O_x, L_z)$  &    &   &   &   &   &      \\
$O^b_{1\pm}/O^c_{1\pm}, (O^b_{2-}, O^c_{2-}) $ &  **\ \,   $\surd$        & **\ \, $\surd$  & $\surd$  &   $\surd$     & $\surd$  &  X     \\
 $(O_x, T_x)$  &    &   &   &   &   &       \\
\hline\hline
\end{tabular}
\vskip -0.05cm
\end{table*}

\vskip 0.5cm
As in the \textbf{($2 + 2$) case} discussed in preceding Sec.~\ref{sec:22}, here we have also the restriction of no equal relative-phase magnitudes in order to enable the sets of two pairs of observables, as given in Table.~\ref{tab:211}, to resolve the phase ambiguity.  Analogous considerations for the \textbf{($2 + 2$) case} allows us to identify the additional observables required to resolve the phase ambiguity when this restriction is not met. They are indicated also in Table.~\ref{tab:211} for the \textbf{case} $2 (a) + 1 (b) + 1 (c)$.

\section{ Phase fixing for the ($3+ 1$) and 4 cases} \label{sec:314}

It is straightforward to show that no sets of observables with the \textbf{($3 + 1$)} or \textbf{($4$) cases} can resolve the phase ambiguity.

Consider the  \textbf{($3+1$) case} of three observables from, say, group $a={\cal BT}$ and one from group $b={\cal BR}$.  Then, from Eqs.(\ref{eq:comb1a},\ref{eq:comb1b}), we have the following possible sets of four observables:
$[(O^a_{n\nu}, O^a_{n'\nu'}), (O^a_{n''\nu''}, O^b_{n'''\nu'''})]$, with $[n,n',n'',n'''=1,2; \nu,\nu',\nu'',\nu'''=\pm\ {\rm and}\ (n,\nu) \ne (n',\nu')\ {\rm and}\ (n'',\nu'') \ne (n,\nu), (n',\nu')]$. For concreteness, consider the set $[(O^a_{1+}, O^a_{1-}), (O^a_{2+}, O^b_{1+})]$. The
pair of observables $(O^a_{1+}, O^a_{1-})$ determines $\sin\phi_{13}$ and $\sin\phi_{24}$ uniquely, yielding the two-fold ambiguity for each of the relative phases $\phi_{13}$ and $\phi_{24}$ as given by Eq.(\ref{eq:2-2}).  This, then, leads to the following four possible expressions for the observable $O^a_{2+}$: 
\begin{align}
O^a_{2+} & = B_{13}\cos\phi_{13} + B_{24}\cos\phi_{24} \nonumber \\
& = \left\{
\begin{array}{l}
\ \ \ B_{13}\cos\alpha_{13} + B_{24}\cos\alpha_{24} \ , \\
\ \ \ B_{13}\cos\alpha_{13} - B_{24}\cos\alpha_{24} \ , \\
- (B_{13}\cos\alpha_{13} + B_{24}\cos\alpha_{24}) \ , \\
- (B_{13}\cos\alpha_{13} - B_{24}\cos\alpha_{24}) \ ,
\end{array} \right.
\label{eq:4-1}
\end{align}
where Eq.(\ref{eq:sclambda}) has been used. Since these expressions are all linearly independent, only one of them will be satisfied - except perhaps for a few special cases - once $O^a_{2+}$ is measured. That is, $O^a_{2+}$ should in principle be able to resolve the discrete ambiguities of $\phi_{13}$ and $\phi_{24}$. The remaining observable $O^b_{1+}$,
\begin{equation}
O^b_{1+} = B_{14}\sin\phi_{14} + B_{23}\sin\phi_{23} \ ,
\label{eq:4-2}
\end{equation}
however, can determine neither $\phi_{14}$ nor $\phi_{23}$, one of which is needed, in addition to $\phi_{13}$ and $\phi_{24}$, for resolving the phase ambiguity of the transversity amplitude up to an arbitrary phase. The analogous reasoning applies to all other sets of four observables in the \textbf{( $3+1$) case}. The reader may convince himself/herself that none of these sets are capable of resolving the phase ambiguity.

\vskip 0.5cm
In the case of four observables from one given group  \textbf{($4$) case}, say,  $[(O^a_{1+}, O^a_{1-}), (O^a_{2+}, O^a_{2-})]$, it is clear from Eq.(\ref{eq:comb1a}) that  they determine the relative phases $\phi_{13}$ and $\phi_{24}$ uniquely, but no information about a third relative phase is available for resolving the phase ambiguity.

\section{Identifying when the equal relative-phase-magnitudes condition  occurs}\label{sec:equalphase}

As we have seen in Secs.~\ref{sec:22} and \ref{sec:211}, the completeness condition for a set of four  double-spin observables to resolve the phase ambiguity of the transversity amplitude holds, provided the equal relative-phase-magnitudes relation (cf.~Eq.(\ref{eq:3-0})) is not met.
This restriction  wouldn't cause a significant problem if this is a rarely occurring  situation.  However, we find no reason \textit{a priori} to expect that this is indeed a rare case. This forces us to verify if the no equal relative-phase condition is met    
for each kinematics (total energy of the system and meson production angle) where the four double-spin observables are measured, for the completeness argument that only four carefully selected double-spin observables are needed. 
Can we know when the equal-magnitudes relation are realized?  The answer to this question is yes as we show in the following.

To be concrete, consider  the pair of observables of the form $(O^a_{n\pm}, O^a_{n\mp})\ (n=1,2)$, from Eqs.(\ref{eq:2-1},\ref{eq:2-2},\ref{eq:2-4}).  When the corresponding phases satisfy $\alpha_{13} = \pm \alpha_{24}$, these observables obey the relation
\begin{equation}
B_{13} \left(O^a_{n \pm} - O^a_{n\mp} \right) = \pm B_{24} \left(O^a_{n \pm} + O^a_{n\mp} \right) \ .
\label{eq:3-1}
\end{equation}
Hence, by measuring the cross section and single-spin observables (which determine $B_{13}$ and $B_{24}$) and the double-spin observables in the above equation, we will be able to gauge if the equal magnitudes relation, $|\alpha_{13}| = |\alpha_{24}|$, is met. Note that in the particular case of $\alpha_{13}=\alpha_{24}=0$, we have 
\begin{equation}
O^a_{1+}  = O^a_{1-} = 0  \quad {\rm and} \quad 
\frac{O^a_{2-}}{O^a_{2+}}  = \frac{B_{13} - B_{24}}{B_{13} + B_{24}} \ .
\label{eq:3-1a}
\end{equation}

\vskip 0.5cm
For the pair of observables of the form $(O^a_{1\pm}, O^a_{2\pm})$ ($\pm$ signs are independent),  from 
Eq.(\ref{eq:2-5}), when $\alpha_{13} = \pm \alpha_{24}$, we have
\begin{equation}
O^a_{1\pm}{^2} + O^a_{2\pm}{^2} = (B_{13} \mp B_{24})^2 \ .
\label{eq:3-2a}
\end{equation}
Note that the $\pm$ sign on the right-hand-side of the above equation goes with  the $\pm$ sign of $\alpha_{24}$.
In the particular case of $\alpha_{13}=\alpha_{24}=0$, we have 
\begin{equation}
O^a_{1\pm} = O^a_{2\pm}=0 \quad {\rm and}\quad B_{13} = \pm B_{24} \ .
\label{eq:3-2aa}
\end{equation}

\vskip 0.5cm
For the pair $(O^b_{1-}, O^c_{1+})$, when $\alpha^{\lambda\lambda'}_{12}(\eta)  = \pm \alpha^{\lambda\lambda'}_{34}(\eta)$, we have, from Eqs.(\ref{eq:12LL'},\ref{eq:34}),
\begin{equation}
\frac{A^{\lambda\lambda'}_c O^b_{1-} + \eta\, A_s \sqrt{D^{\lambda\lambda'\, 2} - \left(O^b_{1-}\right){^2}}}{D^{\lambda\lambda'\, 2}} = \frac{O^c_{1+}}{B_{12} \pm B_{34}} \ ,
\label{eq:3-3a}
\end{equation}
where $A^{\lambda\lambda'}_c$, $A_s$ and $D^{\lambda\lambda'\, 2}$ are given by Eqs.(\ref{eq:AcAs1},\ref{eq:12LL'}).
In the particular case of $\alpha^{\lambda\lambda'}_{12}(\eta)  = \alpha^{\lambda\lambda'}_{34}(\eta) = 0$,
we have 
\begin{equation}
O^c_{1+} = 0 \quad {\rm and} \quad |O^b_{1-}| = |A_s|  \ ,
\label{eq:3-3aa}
\end{equation}
where Eq.(\ref{eq:D}) has been also used.
Equations (\ref{eq:3-3a},\ref{eq:3-3aa}) hold for all the pairs of observables of the form $(O^b_{n\pm}, O^c_{n\pm})\ (n=1,2$ and $\pm$ signs are independent) with the appropriate signs of $B_{23}$ and $B_{34}$ in $A^{\lambda\lambda'}_c$, $A_s$ and $D^{\lambda\lambda'\, 2}$, and also of $B_{12}$ and $B_{34}$.

\vskip 0.5cm
Analogously, for the pair $(O^b_{2-}, O^c_{1+})$, from Eqs.(\ref{eq:27},\ref{eq:34a}), we obtain when $\alpha^{\lambda\lambda'}_{12}(\eta)  = \pm \alpha^{\lambda\lambda'}_{34}(\eta)$,
\begin{equation}
\frac{- A^{\lambda\lambda'}_s O^b_{2-} + \eta\, A_c \sqrt{D^{\lambda\lambda'\, 2} - \left(O^b_{2-}\right){^2}}}{D^{\lambda\lambda'\, 2}} =  \frac{O^c_{1+}}{B_{12} \pm B_{34}} \ ,
\label{eq:3-3b}
\end{equation}
where $A^{\lambda\lambda'}_s$, $A_c$ and $D^{\lambda\lambda'\, 2}$ are given by Eqs.(\ref{eq:AcAs3},\ref{eq:25a}).
In the particular case of $\alpha^{\lambda\lambda'}_{12}(\eta)  = \alpha^{\lambda\lambda'}_{34}(\eta) = 0$,
we have 
\begin{equation}
O^c_{1+} = 0 \quad {\rm and} \quad |O^b_{2-}| = |A_c| \ .
\label{eq:3-3bb}
\end{equation}
Equations (\ref{eq:3-3b},\ref{eq:3-3bb}) hold for all the pairs of observables of the form $(O^b_{2\pm}, O^c_{1\pm})$ ( $\pm$ signs are independent) with the appropriate signs of $B_{23}$ and $B_{34}$ in $A^{\lambda\lambda'}_c$, $A_s$ and $D^{\lambda\lambda'\, 2}$, and also of $B_{12}$ and $B_{34}$.

\vskip 0.5cm
Equations (\ref{eq:3-1},\ref{eq:3-1a},\ref{eq:3-2a},\ref{eq:3-2aa},\ref{eq:3-3a},\ref{eq:3-3aa},\ref{eq:3-3b},\ref{eq:3-3bb}) enable us to gauge when the equal relative phase  magnitudes relation is met for any of the sets of two pairs of observables as listed in Tables.~\ref{tab:bb}, \ref{tab:ab} and \ref{tab:211}, which - otherwise - can resolve the phase ambiguity.

\section{Summary}\label{sec:summary}

By revealing and exploiting the underlying symmetries of the relative phases of the pseudoscalar photoproduction amplitude, we have provided a consistent and explicit mathematical derivation of the completeness condition for the observables in this reaction covering all the relevant cases. In particular, we have determine all the possible sets of four observables that resolve the phase ambiguity of the transversity amplitude up to an overall phase. 
The present work  substantiates and corroborates the original findings of Ref.\cite{ChT97}. However, the completeness condition of a set of four double-spin observables to resolve the phase ambiguity holds only if the relative phases do not have equal magnitudes as specified in Eq.(\ref{eq:3-0}). 
In situations where the equal-magnitudes condition occur, we have shown that one or two or even three extra chosen observables are required, depending on the particular set of two pairs of observables considered as given in Tables.~\ref{tab:bb}, \ref{tab:ab} and \ref{tab:211}, resulting in five or six or seven as the minimum number of chosen double-spin observables required to resolve the phase ambiguity.
In the particular case of vanishing relative phases, we need, eight chosen observables to resolve the phase ambiguity.
This results in a minimum of up to twelve chosen observables to determine the amplitude up to an overall phase: four, to determine the magnitudes of the basic four transversity amplitudes that comprise the full photoproduction amplitude and, up to eight more to resolve the phase ambiguity depending on the particular set of four double-spin observables.

To apply the argument of the completeness condition of a set of four double-spin observables to resolve the phase ambiguity of the photoproduction amplitude, 
 we need to make sure that the restriction of no equal relative-phase magnitudes, as specified in Eq.(\ref{eq:3-0}), is satisfied.
We have shown that it is possible to gauge whether this restriction is satisfied or not for each kinematics where the set of four double-spin observables is measured, because, these observables obey the well defined relationships that are unique to the case of equal relative-phase magnitudes, as seen in Sec.\ref{sec:equalphase}.

\vskip 0.3cm
We  also remark  that quantum mechanics does not allow us to determine the overall phase of the reaction amplitude from experiment. For this, some physics input is required. This fact must have a strong impact on partial-wave analysis in the context of complete experiments for extracting the baryon resonances since, if the overall phase of the amplitude is unknown, the corresponding partial-wave amplitude is an ill defined quantity.   
The issues related to the unknown overall phase have been discussed earlier by several authors. In particular,
Omelaenko \cite{Om81} mentioned the overall phase problem for photoproduction in the summary section of
his paper on discrete ambiguities in truncated partial-wave analysis. In the classic review paper by Bowcock and Burkhardt \cite{BB75}, this problem is discussed as well. 
Dean and Lee \cite{DL72} also investigated this problem mainly for the formalism of $\pi N$-scattering. Two recent publications \cite{WSWTB17,SWOHOSKNOTW18} treat the same problem, but mostly in the simpler context of spinless particle scattering. 

\vskip 0.3cm
Finally, the present type of analysis may be applied to other reaction processes where the interest in determining the complete experiments exist.

\section*{Acknowledgment}

The author is indebted to Frank Tabakin and Harry Lee for many invaluable discussions  and for the encouragement to publish the present results. The author is also grateful to Frank Tabakin for a careful reading of the manuscript.

\end{document}